\pdfoutput=1
\RequirePackage{ifpdf}
\ifpdf 
\documentclass[pdftex]{sigma}
\else
\documentclass{sigma}
\fi

\newcommand{\half}{\frac{1}{2}}
\newcommand{\im}{{\rm i}}
\newcommand{\dd}{\mathrm{d}}
\newcommand{\SL}{\mathrm{SL}(2,\mathbb{C})}
\newcommand{\SU}{\mathrm{SU}(2)}
\newcommand{\Spin}{\mathrm{Spin}(4)}
\newcommand{\SO}{\mathrm{SO}(4)}
\newcommand{\hphi}{\hat{\varphi}}
\newcommand{\hphid}{\hat{\varphi}^{\dagger}}
\newcommand{\ket}[1]{\left| #1 \right\rangle}
\newcommand{\bra}[1]{\left\langle #1 \right|}
\newcommand{\braket}[2]{\langle #1 | #2 \rangle}
\newcommand{\iu}{\mathrm{i}}
\newcommand{\ee}{\mathrm{e}}

\numberwithin{equation}{section}

\begin{document}

\allowdisplaybreaks

\renewcommand{\thefootnote}{$\star$}

\newcommand{\arXivNumber}{1602.08104}

\renewcommand{\PaperNumber}{082}

\FirstPageHeading

\ShortArticleName{Quantum Cosmology from Group Field Theory Condensates: a Review}

\ArticleName{Quantum Cosmology from Group Field Theory\\ Condensates: a Review\footnote{This paper is a~contribution to the Special Issue on Tensor Models, Formalism and Applications. The full collection is available at \href{http://www.emis.de/journals/SIGMA/Tensor_Models.html}{http://www.emis.de/journals/SIGMA/Tensor\_{}Models.html}}}

\Author{Steffen GIELEN~$^\dag$ and Lorenzo SINDONI~$^\ddag$}

\AuthorNameForHeading{S.~Gielen and L.~Sindoni}

\Address{$^\dag$~Theoretical Physics, Blackett Laboratory, Imperial College London, London SW7 2AZ, UK}
\EmailD{\href{mailto:s.gielen@imperial.ac.uk}{s.gielen@imperial.ac.uk}}

\Address{$^\ddag$~Max Planck Institute for Gravitational Physics, Am M\"uhlenberg 1, 14476 Golm, Germany}
\EmailD{\href{mailto:lorenzo.sindoni@gmail.com}{lorenzo.sindoni@gmail.com}}

\ArticleDates{Received February 29, 2016, in f\/inal form August 12, 2016; Published online August 18, 2016}

\Abstract{We give, in some detail, a critical overview over recent work towards deriving a~cosmological phenomenology from the fundamental quantum dynamics of group f\/ield theory (GFT), based on the picture of a macroscopic universe as a~``condensate'' of a large number of quanta of geometry which are given by excitations of the GFT f\/ield over a~``no-space'' vacuum. We emphasise conceptual foundations, relations to other research programmes in GFT and the wider context of loop quantum gravity (LQG), and connections to the quantum physics of real Bose--Einstein condensates. We show how to extract an ef\/fective dynamics for GFT condensates from the microscopic GFT physics, and how to compare it with predictions of more conventional quantum cosmology models, in par\-ti\-cular loop quantum cosmology (LQC). No detailed familiarity with the GFT formalism is assumed.}


\Classification{83C45; 83C27; 83F05; 81T27}

\renewcommand{\thefootnote}{\arabic{footnote}}
\setcounter{footnote}{0}

\section{Introduction}

The idea that spacetime and geometry should be emergent, i.e., an ef\/fective description of the collective behaviour of dif\/ferent, ``pre-geometric'' fundamental degrees of freedom, is now a common theme in most approaches to quantum gravity \cite{Danielephil}. In particular, in theories formulated fundamentally in terms of discrete, combinatorial structures rather than continuum variables (such as a spacetime metric or connection on a smooth manifold) in order to avoid the problem of perturbative non-renormalisability of general relativity \cite{nonrenorm}, the emergent nature of a~continuum description is a~necessity, in order to connect to other areas of physics and to our experience which take place in the continuum. Showing how a continuum emerges is hence the challenge faced by loop quantum gravity \cite{LQGrev,LQGbook1,LQGbook2} and spin foam models \cite{SFrev,SFlectures}, causal dynamical triangulations~\cite{CDTreview}, causal set theory~\cite{CSTreview}, as well as tensor models \cite{tensorreview} and group f\/ield theories~\cite{GFTrev1,GFTrev2,GFTrev3}, a~challenge which has been addressed with varying success. A~se\-pa\-rate notion of emergence of dynamical spacetime arises in the AdS/CFT correspondence~\cite{ADSCFT} for supersymmetric theories, which posits that a continuum quantum spacetime geometry is in a precise sense dual to a lower-dimensional conformal f\/ield theory on a f\/ixed geometry. These rather dif\/ferent notions of emergence all face the challenges of (nonlocally~\cite{Marolf}) mapping the fundamental degrees of freedom to the sought-after emergent spacetime description.

In a group f\/ield theory (GFT), spacetime emerges from combinatorial structures, labelled by additional geometric or pre-geometric data. In the simplest representation of simple GFT models, this data consists of group elements attached to the edges of a graph, typically of f\/ixed valency. The group elements themselves are interpreted as parallel transports of a gravitational connection along these edges, and hence are analogous to the variables of lattice gauge theory. Unlike in lattice gauge theory, the graph or lattice does not carry a clear physical interpretation: graphs appear as Feynman diagrams in the perturbative expansion around a GFT vacuum, and quantities of interest such as the partition function contain a summation over an inf\/inite number of such graphs. In general, there is little reason to trust perturbation theory, i.e., a~truncation of this sum to a few simple graphs, since there is no small expansion parameter and no interpretation of the vacuum as a physically meaningful geometry. Indeed, viewing GFT as a~second quantisation formulation of loop quantum gravity (LQG) and hence adopting operators for geometric observables such as volumes and areas from LQG~\cite{Daniele2ndq}, the most natural GFT vacuum is a state in which expectation values for such geometric observables vanish~-- it is a~`no-space' state, analogous to the Ashtekar--Lewandowski vacuum~\cite{ALvacuum} of LQG. An approximately smooth, macroscopic geometry of physical interest can hardly be found as a small perturbation of such a vacuum.

Since the early days of LQG, some ef\/fort has been invested into the construction of states that can approximate continuum geometries, such as f\/lat space \cite{weave}. These necessarily excite a large number of degrees of freedom compared to the Ashtekar--Lewandowski vacuum, and accordingly are described by a large graph with many edges and vertices. One would hope to f\/ind states that can play the role of a new vacuum of the theory around which a dif\/ferent notion of perturbation theory can be def\/ined, e.g., in order to compare LQG with the perturbative quantisation programme of continuum general relativity. Completing this task requires knowledge about the dynamics of many interacting degrees of freedom, which has been dif\/f\/icult to extract from traditional LQG and its spin foam formulation, partly due to the rather complicated combinatorial structure of existing models. In the study of spin foam models such as the currently favoured EPRL model~\cite{EPRL}, known results for transition amplitudes are mainly restricted to asymptotic regimes, in particular in the ``large-$j$ limit'', see, e.g.,~\cite{nottingham,muxin}.

The situation calls for the development of systematic methods to treat many degrees of freedom in LQG, beyond the truncation given by a f\/ixed graph. One possibility is to adopt methods from lattice gauge theory and use coarse graining or renormalisation procedures in order to obtain solutions to the dynamics and ultimately a theory with a well-def\/ined continuum limit, where dif\/feomorphism symmetry is restored \cite{Biancarev}. The endpoint of such a renormalisation group f\/low corresponds to a particular continuum phase, describing the collective behaviour of many degrees of freedom (inf\/initely many in the limit). In an interpretation of the setting as a theory of quantum gravity, one would like to show that such a phase describes a smooth non-degenerate spacetime geometry, such as Minkowski or de Sitter space.

\looseness=-1 In the GFT setting, one can also explore a dif\/ferent route. As a second quantisation formulation of LQG, the main advantage of GFT is the availability of quantum f\/ield theory methods when treating many degrees of freedom in the language of quantum f\/ields rather than $N$-particle states. Conceptually, the situation is rather similar to the use of (non-relativistic) quantum f\/ield theory in condensed matter physics, when one is interested in the collective behaviour of a large number of interacting atoms or molecules, rather than perturbative scattering around the Fock vacuum. One of the most relevant examples of the successful application of quantum f\/ield theory in condensed matter physics is the study of Bose--Einstein condensation and superf\/luidity~\cite{BECbook}. In this context, the condensation of many atoms into a~common ground state can be viewed as a~transition from a perturbative phase around the Fock vacuum (of zero atoms) into a condensed phase, with associated symmetry breaking of the ${\rm U}(1)$ symmetry of the theory. Similarly, the language of GFT allows a treatment of many degrees of freedom through quantum f\/ields rather than a f\/ixed graph or ``$N$-particle state''. The elementary building block of a graph, a vertex of given valency, then plays the role of an {\em atom of geometry}. In fact, this interpretation is quite lite\-ral\-ly the one in LQG: the volume operator only has non-zero eigenvalues when acting on vertices~\cite{LQGbook1,LQGbook2}, so that such vertices do indeed form the ``atoms of space''. Using not a f\/ixed number of such atoms but more general states that contain a superposition of many (typically inf\/initely many) graphs is quite natural in LQG where the (kinematical) Hilbert space is a direct sum of Hilbert spaces associated to dif\/ferent graphs~\cite{LQGbook1,LQGbook2}. The continuum limit of an inf\/inite number of such atoms is the usual thermodynamic limit of statistical f\/ield theory. This view suggests that a very natural candidate for a conf\/iguration of many degrees of freedom with high coherence and symmetry, and potentially a candidate for a non-perturbative vacuum of LQG, should be given by a {\em group field theory condensate} analogous to a Bose--Einstein condensate~\cite{Danielecond1,Danielecond2}.

Given that this idea of a macroscopic spacetime as a kind of condensate of microscopic degrees of freedom is rather natural, it may not appear surprising that it has already been suggested in other contexts. In LQG, it was suggested by Koslowski \cite{koslowski} and shown in more detail by Sahlmann \cite{sahlmann} that the Ashtekar--Lewandowski construction, leading to a unique vacuum in LQG \cite{LOST}, could be generalised to allow for a larger class of vacua, where expectation values for geometric operators are shifted by a given classical f\/ield conf\/iguration with respect to the Ashtekar--Lewandowski vacuum, so that one can choose for example a vacuum corresponding to continuum f\/lat space. One can view the Koslowski--Sahlmann vacua as implementing a breaking of dif\/feomorphism symmetry which would require these expectation values to vanish. There has not yet been much work on understanding these vacua as a dynamically emerging conf\/iguration; excitations over them are also still discrete structures as in usual LQG.

There is also a more ``bottom-up'' motivation from quantum cosmology for thinking of a~macroscopic universe as a~sort of condensate \cite{QCReview}. In quantum cosmology, the traditional stra\-te\-gy since Misner~\cite{Misner} for obtaining a manageable quantum model of the universe has been to perform a ``minisuperspace'' reduction at the classical level, i.e., to assume that the universe is spatially homogeneous, and to quantise only the remaining degrees of freedom of geometry. The assumption of exact homogeneity means that one focuses on the local evolution of a small patch around a single point in space, as there are no spatial gradients by construction, and one imposes the symmetry-reduced quantum Einstein equations as the dynamics of such a patch. As explained in~\cite{QCReview}, this procedure can be thought of as a~``single-patch'' quantum model in which one quantises a single ``atom'' or chunk of space, and assumes that the dynamics of the atom is related to the ef\/fective dynamics at large scales. This view would suggest passing to a~``many-patch model'' of many interacting chunks of space, as a way to check the validity of this assumption, and to extend the model to include inhomogeneities. Depending on the length scales associated to these chunks, this idea can be related to the separate universe picture in cosmology~\cite{sepuniv}, where inhomogeneities in a certain range of wavelengths can be included by modelling an inhomogeneous universe as composed of many small, locally homogeneous patches. In such a many-patch model, given that inhomogeneities are very small in the observed Universe, one then considers states with weak interactions between dif\/ferent patches. A simple ansatz for the many-patch wavefunction would be a product state, so that there is a collective single-particle wavefunction describing the state, again very similar to states used for (weakly interacting) Bose--Einstein condensates. This idea was explored in~\cite{NonlinQC} in a lattice model starting from classical general relativity, where one f\/inds that the existence of interactions between patches leads to a nonlinear and nonlocal evolution equation for the single-particle wavefunction, similar to the Gross--Pitaevskii equation in condensed matter physics~\cite{BECbook}. We will f\/ind very similar qualitative features in the ef\/fective dynamics of GFT condensates.

The application of loop quantum gravity techniques to quantum cosmology led to the development of loop quantum cosmology (LQC)~\cite{LQCrev1,LQCrev2}. LQC combines ``top-down'' and ``bottom-up'' elements: while the classical Friedmann equation is used as a starting point for the dynamics, one also uses the kinematics of LQG, so that the continuum Ashtekar--Barbero connection \cite{Ashtekar,Barbero} is replaced by f\/inite parallel transports along edges of a graph, and the Wheeler--DeWitt equation becomes a dif\/ference rather than a dif\/ferential equation. An open issue that has attracted a lot of interest in recent years (see, e.g., \cite{LQCLQG1,LQCLQG2,LQCLQG3,LQCLQG4,LQCLQG5} for various approaches) is how to embed the kinematics and dynamics of LQC into full LQG, or rather derive LQC as an approximation or ef\/fective description of the full theory of LQG when applied to almost exactly homogeneous, macroscopic Universes like our own. In this review, we explore another viewpoint in which one thinks of quantum cosmology as the ef\/fective dynamics of a {\em condensate} of microscopic degrees of freedom of quantum geometry. In short, ``{\em quantum cosmology is the hydrodynamics of quantum gravity}''. The condensate hypothesis provides methods for deriving an ef\/fective description of the dynamics of realistic universes in GFT, which can then be compared with conventional quantum cosmology models such as LQC, as well as ultimately with observation.

\subsection{Relation to analogue and emergent gravity}

The philosophy we are following shares some common ground with the philosophy behind the programmes of ``analogue'' and ``emergent'' gravity. Analogue models for gravity \cite{AnalogueReview} seek to emulate some physical properties of gravitational systems, such as Hawking radiation \cite{unruh}, by non-gravitational systems usually described by some form of hydrodynamics. Motivated by the derivation of the Einstein equations from thermodynamics \cite{jacobson}, and more generally the close link between spacetime geometry and thermodynamics, following Boltzmann's dictum ``if you can heat it, it must have microstructure'' \cite{padmanabhan}, emergent models go one step further and aim to describe {\em dynamical} gravity as emerging from an ef\/fective description of dif\/ferent fundamental degrees of freedom \cite{emergentreview}. The idea that spacetime is a condensate of fundamental ``atoms of spacetime'' has been suggested in this context of emergent gravity \cite{hucondens}, without an obvious proposal for the nature of the ``atoms''.

Group f\/ield theories provide a natural candidate for a theory of the atoms of spacetime and their dynamics, and suggest that classical gravitational dynamics can be found in an ef\/fective hydrodynamic description of the dynamics of a GFT condensate, as suggested in the context of lower-dimensional models in~\cite{DanieleLorenzoHydro}. The results and methods covered in this review show the detailed application of these ideas to cosmology.

\subsection{Relation to studies of renormalisation and phase structure in LQG/GFT}\label{renormsec}

The study of condensates in GFT can be seen as part of a larger ef\/fort aimed at under\-standing the collective dynamics of many interacting degrees of freedom of quantum geometry in loop quantum gravity and its generalisations, including the more general setting of GFT. Given the interpretation of GFT condensates as macroscopic continuum geometries, one would like to show that such condensates correspond to a dynamical phase of a known GFT model, at least in a~certain range of the free parameters (coupling constants). One would also like to understand the possible phase transitions, in particular transitions between a ``pre-geometric'' and a~``geometric'' continuum phase such as described by a condensate. Replacing the classical Big Bang singularity by such a phase transition is the idea often dubbed {\em geometrogenesis}, and f\/irst advocated in the context of ``quantum graphity''~\cite{graphity}. Showing that geometrogenesis can occur in the setting of GFT would mean a~revolutionary reinterpretation of the origin of space and time, as it would show the emergence of a~GFT phase with macroscopic geometric interpretation (in terms of a~four-dimensional, continuum spacetime, causal structure, a metric, etc.) from a~framework which is fundamentally formulated without any of these structures, and instead just uses combinatorial and group-theoretic data.

The main tool for mapping out the phase diagram for GFT and related theories is the renormalisation group, particularly its Wilsonian formulation based on ef\/fective actions. Applying renormalisation tools to GFT has been an extremely active f\/ield of research in recent years, see, e.g., \cite{GFTrenorm1,GFTrenorm2,GFTrenorm3,GFTrenorm4,GFTrenorm5}. A major result is that many models studied so far (of the {\em tensorial} GFT type) exhibit asymptotic freedom \cite{GFTrenorm1,vincentletter}, making them potential candidates for a~fundamental theory. The infrared limit of the renormalisation f\/low is then important in order to understand the hypothesis that spacetime is a GFT condensate (of many ``small'' quanta). Here, in some models evidence has accumulated for phase transitions between a symmetric and a condensate-type phase \cite{GFTphasetrans, GFTphasetrans2}, just of the type we are interested in, with an associated f\/low to an interacting f\/ixed point in the IR. So far, no such renormalisation analysis has been performed for GFT models with direct relation to LQG, such as the EPRL/FK GFT \cite{EPRLGFT}, but progress towards the study of more and more sophisticated models has been rapid. Ultimately, one would aim to justify the assumption of a GFT condensate in the models we are looking at from a more rigorous analysis of this type.

There has also been a lot of activity in the renormalisation analysis of simplif\/ied and gene\-ralised spin foam models by coarse graining methods, aimed at f\/inding a continuum limit that restores dif\/feomorphism invariance \cite{bianca1,bianca2,bianca3}. Part of this ef\/fort has culminated in the development of a new representation for LQG \cite{BenniBiancaMarc,BiancaMarc}, with a new vacuum that is in some sense dual to the Ashtekar--Lewandowski vacuum of traditional LQG: it is a vacuum of exactly f\/lat connections but completely undetermined conjugate metric variables. These results already show that phases of LQG and spin foam models beyond the degenerate phase around the Ashtekar--Lewandowski vacuum exist, giving further motivation to the study of GFT condensates, although the detailed connection between these research programmes still needs to be understood better.

\subsection{Structure of this review}

After giving some perspective in this introductory section, our aim is to give a self-contained overview over the research of roughly the last three years on condensates in GFT and their relation to quantum cosmology. To this end, in Section~\ref{gftsec} we introduce the basics of the GFT formalism, set up notation for this review, and def\/ine the class of models we will be looking at. While some of this material will be known to readers familiar with the GFT literature, they might want to consult it for clarity on which GFT models we are studying and on notational conventions. In Section~\ref{kinesec}, we outline the (kinematical) def\/inition of condensate states in GFT, mainly focusing on the simplest case that is in direct analogy with real Bose--Einstein condensates. Many conceptual and technical points regarding the interpretation of these condensates in terms of spacetime geometry will be explained. We also touch upon a more general class of condensate states introduced more recently. In Section~\ref{dynsec}, we show how to extract information about the dynamics of such condensates from the fundamental quantum GFT. We give dif\/ferent methods for extracting ef\/fective dynamical equations, explain the relation to quantum cosmology and the approximations and truncations involved. We also go into more detail as to how such ef\/fective dynamics can be compared with models such as LQC. In Section~\ref{resultsec}, we f\/inally summarise the main results, open questions and future directions of this approach.

\section{Group f\/ield theory (GFT)}\label{gftsec}

In this section we will present the main ideas and basic technical tools of GFT that will be needed for the application to cosmology. The presentation will be self-contained; for further background on the general setting of GFT and recent progress, we refer to the reviews \cite{GFTrev1,GFTrev2,GFTrev3}.

{\em Group field theories} are quantum (or statistical) f\/ield theories def\/ined over group manifolds, most commonly restricted to classes of theories which can be designed to provide a possible def\/inition of a path integral for quantum gravity. In a nutshell, we are interested in models for which the perturbative expansion of its partition function or f\/ield equation, and the perturbative expansion of the correlation functions, generates a sum over graphs that can be seen as dual to discretisations of manifolds.

In addition to the peculiar combinatorial structure, similar to that found in matrix and tensor models after which GFTs are modelled, the graphs of GFT are decorated by data related to the chosen group (group elements, representation labels or Lie algebra elements). The role of these additional data is to store geometric information (metric or connection data) beyond the mere combinatorial/topological structure. In fact, this group-theoretic data corresponds precisely to the basic variables of LQG, holonomies of a connection and f\/luxes of a triad f\/ield. In this way, GFT attempts to def\/ine a sum over discretised geometries which can be used, once a continuum limit is identif\/ied, to obtain a path integral for quantum gravity.

The def\/initions given so far cover a fairly rich class of models. In this review, we focus on simple models that have so far mainly been used in the application to cosmology, but the constructions of GFT condensate states have straightforward extensions to more complicated models. In particular, the desired property of the sum over graphs -- that it represents a sum over combinatorial manifolds, or at least pseudomanifolds -- is generally achieved by introducing {\em coloured} models with more than one fundamental GFT f\/ield \cite{colourGFT,GurauRyan}. We will not be discussing coloured models in any detail, but it should be clear in the following that the main ideas are generally applicable to a wide class of GFT models.

\subsection{Def\/inition of the basic models}

There are two prototypes of GFT that are useful for applications to quantum gravity, which both def\/ine a quantisation of topological BF theories over random manifolds. The f\/irst one is the Boulatov model \cite{Boulatov}, whose partition function can be naturally rewritten in
terms of the Ponzano--Regge model \cite{PonzanoRegge} for three-dimensional (Riemannian) quantum gravity.

The second model, def\/ined by Ooguri \cite{Ooguri}, can be seen as the four-dimensional counterpart of the Boulatov model. It generates a Feynman expansion which can be related to the amplitudes of a topological BF quantum f\/ield theory, the backbone of the spin foam approach to quantum gravity in four dimensions \cite{SFrev}. For the application to cosmology, we focus on four dimensions, so that we will be interested in models built on extending the Ooguri model from a topological theory to a theory of quantum gravity. We will comment later on various possible generalisations. We will also focus on the pure gravity case, as most models so far aim to def\/ine pure quantum gravity. The extension to matter coupling is of course crucial for a more realistic cosmology, and will be discussed in Section~\ref{dynsec}.

\begin{figure}[t]\centering
\includegraphics[width= 4cm]{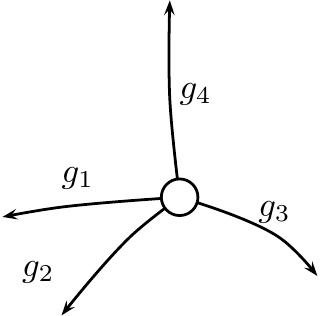}
\qquad
\includegraphics[width = 3.5cm]{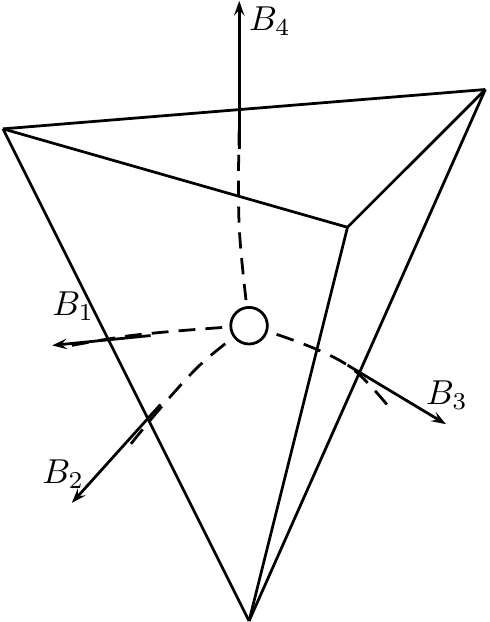}
\caption{GFT quanta as open spin network vertices (left) or quantum tetrahedra (right).}\label{FigGFTquanta}
\end{figure}

The starting point is the choice of a Lie group $G$, which will select the type of data appearing in the Feynman diagrams, and will be interpreted as the local gauge group of gravity. Typical choices are $G=\SU$, $\SO$, $\Spin$ or~$\SL$. The second main choice is the combinatorial structure of the elementary building blocks used to build spacetime geometry. For four-dimensional spacetime and three-dimensional space, the simplest possibility is to use simplicial building blocks: the elementary building block of three-dimensional {\em space} is then a~(quantum) tetrahedron~\cite{Barbieri}. In the dual picture familiar from LQG and spin foams, it corresponds to a central vertex with four open outgoing links, with group-theoretic data associated to these links. The f\/ield theory is accordingly def\/ined on four copies of $G$ (see Fig.~\ref{FigGFTquanta}). With these choices, the elementary quanta of the theory can be assembled into three-dimensional simplicial geometries that represent the boundary states of the theory. This interpretation in terms of boundary conf\/igurations of four-dimensional geometries is then completed by the exa\-mi\-nation of the perturbative expansion of the correlation functions that they encode, once the dynamics with the correct combinatorial structure is introduced. It is the interplay between the combinatorics of the equations of motion and the number of group arguments of the f\/ields, together with the gauge group used, that suggests the interpretation of the quanta and their interactions in geometrical terms. Indeed, with dif\/ferent dynamics one could also construct a~f\/ield theory based on f\/ields with four arguments which has a $2+1$-dimensional interpretation, with basic quanta interpreted in terms of quadrilaterals. Generalisations of GFT in which the restriction to simplicial geometry is dropped and general three-dimensional polyhedra appear as elementary quanta have been constructed~\cite{OritiAllLQG}. These generalisations are motivated primarily by the appearance in canonical loop quantum gravity of graphs with vertices of arbitrary valency, corresponding to general polyhedra~\cite{poly}, where a restriction to tetrahedra seems arbitrary. Restricting to simplicial building blocks means one can keep track of the topology of the graphs transparently, and is certainly the simplest choice, on which we focus in the following.

Superf\/icially, these choices seem to imply that we are assuming spacetime remains four-dimensional up to the very smallest scales, but this is not true: any useful notion of dimensionality would apply to the collective properties of a number of elementary quanta, rather than a~single one. Experience from approaches to quantum gravity that fundamentally involve discrete structures shows that one should then expect a dimensional f\/low when going from macroscopic to microscopic scales, where the nature of this f\/low (and in particular, the question of whether four dimensions are reached at large scales) depends highly non-trivially on the details of the model, and not simply on the ``dimensionality'' used for the fundamental building blocks of the theory. For some results relevant in the GFT context, see \cite{dimflow}.

After these basic choices, the f\/ield theory is specif\/ied by a choice of a type of f\/ield and by the corresponding action. The most common choices are real or complex scalar f\/ields. While topological models can often be based on a real f\/ield, in order to set up the Fock space structure required for the discussion of GFT condensates, a complex f\/ield will be required. In either case, a simple general form of the action, for compact~$G$, is
\begin{gather}\label{simpleaction}
S[\varphi] = \int \dd g\, \dd g'\, \bar\varphi(g_I) K(g,g') \varphi(g_I')
+ \frac{1}{5!}\int (\dd g)^5 V\big(g^{1},\ldots, g^{5}\big) \prod_{j=1}^5 \varphi\big(g_I^{j}\big) + {\rm c.c.},
\end{gather}
where here and in the following we are using a condensed notation. The arguments of the f\/ield are
$g^{i} \equiv g^i_I = (g^i_1,g^i_2,g^i_3,g^i_4)$, and we are often suppressing the lower index running over the four group elements. The symbol $\int \dd g$ stands for an integration over $G^4$, where $\dd g$ is the Haar measure on $G^4$, normalised to unit total volume. Accordingly, $\int (\dd g)^5$ means integration over $G^{4\cdot 5}=G^{20}$. This notation will be useful, as one is almost always interested in $G^4$ or products of several $G^4$ domain spaces. In the few cases where a dif\/ferent integration domain is needed, it will be specif\/ied explicitly (see, e.g., (\ref{identity})), so that no confusion should arise.

One also needs to specify the function space that is assumed to be integrated over in the GFT path integral. In the perturbative approach, we can assume that this space is the space of square-integrable functions on $G^4$. Since the interaction terms involve higher powers of the f\/ields in convolution with variously constructed kernels, it is necessary to check that these terms are well-def\/ined in the given space of functions and, if needed, to introduce a regularisation, as we will discuss in one example below (\ref{fieldsymmetry}).

The kernels $K(g,g')\equiv K(g_I,g'_I)$ and $V(g^{1},\ldots, g^{5})\equiv V(g^1_I,\ldots,g^5_I)$ determine the details of the resulting Feynman amplitudes. The choice of an interaction term containing f\/ive copies of the f\/ield in \eqref{simpleaction} is suggested by the interpretation of these amplitudes in terms of four-dimensional f\/lat simplices glued along common boundaries, i.e., sharing tetrahedra on their boundaries.

This action can be generalised in many dif\/ferent ways. One obvious possibility is to introduce more terms in the interaction potential, admitting an interpretation of simplicial polytopes (i.e., polytopes whose boundaries are made of simplices). For now, we focus on the well-understood simplicial case. More importantly, in the construction of models relevant for quantum gravity rather than topological f\/ield theory, one needs to include the so-called simplicity constraints into the interaction $V$; we will detail this in Section~\ref{sfrelation}.

Apart from the motivation coming from the interpretation as simplicial geometry, one can also consider dif\/ferent possible interaction terms in~(\ref{simpleaction}) from a quantum f\/ield theory perspective, in particular by studying their renormalisability. Indeed, the construction of renormalisable and asymptotically free GFT models has been the focus of recent work, see Section~\ref{renormsec}.

An important property of the GFT f\/ield is a type of discrete gauge invariance, similar to lattice gauge theory: as our group elements will be interpreted in terms of parallel transports of a connection, from a vertex to the endpoint of an edge, we should make sure that only gauge-invariant information is used. A~simple choice is to restrict the f\/ield conf\/igurations to those which are diagonal under the right diagonal action of $G$ on $G^{4}$:
\begin{gather}\label{fieldsymmetry}
\varphi(g_1h,g_2h,g_3h,g_4h) = \varphi(g_1,g_2,g_3,g_4), \qquad \forall\, h \in G.
\end{gather}
While this is the simplest option, there are other possibilities: for instance, one can use general f\/ields on~$G^4$ but include into the potential term a suitable projector that maps functions on~$G^4$ onto functions on~$G^4/G$. In the following we will keep using~$G^4$ as the domain of the f\/ields, even though we will impose gauge invariance directly on the f\/ield. While this means that the parametrisation is redundant, it greatly simplif\/ies the construction of the models.

The symmetry (\ref{fieldsymmetry}) leads to some technical complications if one wants to extend the construction to non-compact $G$, in particular $G=\SL$ as the Lorentz group in four dimensions. Indeed, as one can see easily, for $G=\SL$ the action (\ref{simpleaction}) is ill-def\/ined; because of the symmetry property (\ref{fieldsymmetry}) of the f\/ield, there is a redundant integration over $\SL$ with inf\/inite volume. This issue can be f\/ixed rather easily, either by working with a gauge-f\/ixed f\/ield on $\SL^4/\SL\simeq \SL^3$, or by simply removing redundant integrations in the kinetic and interaction terms, keeping arguments of the f\/ields f\/ixed. We refer to Appendix A of \cite{JHEP} for a detailed discussion of these subtleties (and their resolution) for the Ooguri model, but for the purposes of this review just stress that they pose no substantial dif\/f\/iculty.

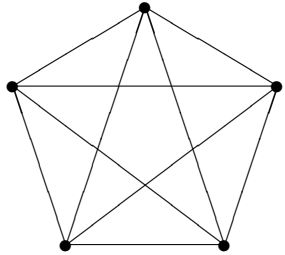
\begin{figure}[t]\centering
\begin{picture}(100,100)
\put(20,0){$\bullet$}\put(80,0){$\bullet$}\put(23,3){\line(1,0){60}}
\put(0,60){$\bullet$}\put(100,60){$\bullet$}\put(50,90){$\bullet$}
\put(23,3){\line(-1,3){20}}\put(83,3){\line(1,3){20}}\put(3,63){\line(5,3){50}}\put(103,63){\line(-5,3){50}}
\put(3,63){\line(1,0){100}}\put(3,63){\line(4,-3){80}}\put(103,63){\line(-4,-3){80}}
\put(23,3){\line(1,3){30}}\put(83,3){\line(-1,3){30}}
\end{picture}
\caption{Complete graph of f\/ive vertices.}\label{graphfig}
\end{figure}

In order to f\/inally be more concrete, choosing a {\em real} f\/ield ($\varphi=\bar\varphi$), the simple choices
\begin{gather*}
K(g,g') = \prod_{I} \delta\big(g'_I (g_I)^{-1} \big)
\end{gather*}
for the kinetic term and
\begin{gather}\label{Ooguri}
V\big(g^{1},\ldots, g^{5}\big)= \lambda \prod_{e} \delta\big(g^{s(e)}_{l(s(e))} \big(g^{t(e)}_{l(t(e))}\big)^{-1}\big)
\end{gather}
for the interaction term lead to the Ooguri model. Here we are denoting by~$e$ the ten edges of a~complete graph of f\/ive vertices (see Fig.~\ref{graphfig}), by~$s(e)$ the copy of the f\/ield associated to the source of the edge, by~$t(e)$ the target, and by $l(s(e))$ (and $l(t(e))$) the group element involved. This choice of combinatorics encodes the pattern of gluings in order to form a 4-simplex out of f\/ive tetrahedra. As we shall see later, by modifying the kernels~$K$ and~$V$ it is possible to generate models that are related in a~precise way to spin foam models~\cite{SFrev}.

The partition function obtained from these ingredients,
\begin{gather*}
Z = \int \mathcal{D} \varphi \exp( -S[\varphi]),
\end{gather*}
when treated as a perturbative expression in the GFT coupling constant~$\lambda$, generates a sum over graphs that can be seen as dual to simplicial complexes. The amplitudes corresponding to the Ooguri model can be shown to be the amplitudes of a discretised BF theory~\cite{Ooguri},
\begin{gather*}
Z = \sum_{\Gamma} \lambda^{\#V(\Gamma)} \mathcal{A}(\Gamma) = \sum_{\Delta(\Gamma)} \lambda^{\#V(\Gamma)} \mathcal{A}_{BF}(\Delta(\Gamma)),
\end{gather*}
with a sum over the graphs $\Gamma$ being reinterpreted as a sum over discrete simplicial comple\-xes~$\Delta(\Gamma)$. It is precisely the possibility of generating a sum over (random) simplicial complexes that makes GFT relevant for the def\/inition of a discrete path integral for the gravitational f\/ield. The choice of the Euclidean partition function assumes implicitly that, after its proper def\/inition, it will be analytically continued, once a suitable parameter has been introduced, to the quantum-mechanical partition function. Since we will be working directly in the quantum operatorial version, this choice plays no role in the rest of the presentation.

The main problem that we are left with, then, is to make sense of formal expressions as this partition function. In this review, we mainly focus on the attempts at the extraction of their physical content which follow the ideas of condensed matter physics for many-body problems.

\subsection{Second quantisation, states associated to closed graphs, and operators}\label{2ndquantsec}

Before discussing the modif\/ication of the topological kernel \eqref{Ooguri} necessary to formulate quantum gravity models and not just topological f\/ield theories, we brief\/ly present GFT from the canonical viewpoint of f\/ield operators. While the operatorial version of the models does not add anything substantially new to the general idea of obtaining spacetime from Feynman diagrams, it is the key ingredient that will allow us to make signif\/icant progress in the calculation of physical predictions. Indeed, the language of Fock spaces and f\/ield operators acting on them will allow us to attack the problem of understanding the physical implications of GFT using a number of powerful techniques adapted from condensed matter physics. Thanks to this reformulation, we will be able to design quantum states with clearer geometric interpretation, which can include some educated guesses about possible nontrivial nonperturbative aspects of the theory in a~simple and controlled manner. Additionally, the connections (and dif\/ferences) to canonical LQG can be stated more precisely~\cite{Daniele2ndq}, without going through the spin foam formalism.

The canonical formalism has clear advantages. Besides the construction of interesting quantum states, in order to answer specif\/ic physical questions we need to understand how to compute the expectation values of physical quantities. The second quantised formulation that we are \mbox{going} to use will allow us to translate and generalise geometric operators introduced in LQG.

The construction follows the standard construction of nonrelativistic quantum f\/ield theories. It is important to stress, however, that we are not viewing it as originating from the canonical quantisation of a f\/ield theory: there is no canonical choice of ``time variable'' on the domain space of GFT that would allow a def\/inition of generalised coordinates and conjugate momenta, in order to justify a particular symplectic structure and corresponding operator algebra. Although it would be possible to def\/ine such an extension of GFT, here we simply postulate the basic commutation relations and resulting Fock space, motivated by the origins of GFT in LQG. Alternative constructions leading to a dif\/ferent Fock quantisation are certainly possible.

The def\/inition of the f\/ield operators involves a choice of statistics for indistinguishable particles. We choose bosonic statistics, compatible with the picture of spacetime as a condensate. The f\/ield operators acting on the Fock space then obey the commutation relations
\begin{gather}\label{commutation}
\big[ \hphi(g),\hphid(g')\big]= \mathbb{I}_R(g,g'),\qquad \big[\hphi(g),\hphi(g')\big]=\big[\hphid(g),\hphid(g')\big]= 0,
\end{gather}
where
\begin{gather}\label{identity}
\mathbb{I}_{R}(g,g') = \int_{G} \dd k \prod_{I=1}^4 \delta\big( g_I k (g'_I)^{-1} \big)
\end{gather}
represents a Dirac delta distribution on the homogeneous space $G^4/G$, ref\/lecting the choice of imposing gauge invariance at the level of the f\/ield~(\ref{fieldsymmetry}). Again, for non-compact groups, (\ref{identity})~needs to be replaced by a regularised expression, corresponding to a Dirac delta on~$G^4/G$.

From the commutation relations (\ref{commutation}), one can now construct a Fock space in the standard way, starting from a Fock vacuum~$\ket{0}$. This vacuum is interpreted as a ``no-space'' state with no quanta, analogous to the Ashtekar--Lewandowski vacuum of LQG. The action of the ladder operators on this state can be naturally interpreted as the creation of a~four-valent open spin network vertex; we are interpreting the state
\begin{gather*}
\ket{g_I} = \hphid(g_I) \ket{0}
\end{gather*}
as a single spin network vertex decorated with the four group elements $g_1,\ldots, g_4$.

States characterised by nontrivial distributions of the group-theoretic data can be obtained straightforwardly from suitable integrations with
wavefunctions,
\begin{gather*}
\ket{f} = \int \dd g\, f(g_I) \hphid(g_I) \ket{0}, \qquad f(g_I) = \braket{g_I}{f} = \bra{0} \hphi(g_I) \ket{f} .
\end{gather*}

This extends directly to multiparticle states. For these, we can add an additional structure to our interpretation of the states, namely a graph. Consider an $N$-particle state
\begin{gather*}
\ket{f_N} = \frac{1}{\sqrt{N!}}\int (\dd g)^N\, f\big(g^1,\ldots, g^N\big) \prod_{j=1}^N \hphid(g^j) \ket{0} .
\end{gather*}
The wavefunction $f(g^1,\ldots, g^N) = \braket{g^1_I,\ldots, g^N_I}{f_N}$ will, in general, depend on all~$4N$ group elements, modulo the vertex gauge invariances from (\ref{fieldsymmetry}). Notice that, by bosonic symmetrisation, the wavefunction is completely symmetric under the permutation of the tuples of group ele\-ments~$g_I$ associated to each vertex. Such generic $N$-particle states are interpreted graphically as states of~$N$ open spin network vertices.

A special class of states, of particular importance in the LQG interpretation, corresponds to the case in which the wavefunction is constructed out of a~generic multiparticle wavefunction with the application of a projection which reduces the dependence of the wavefunction to products of the type $(g_I^i)^{-1} g_J^j$. This is achieved by $2N$ convolutions
\begin{gather}\label{convolution}
f\big(\ldots, g_I^i,\ldots , g_J^j, \ldots\big) = \int_{G^{2N}} \dd h_{IJ}^{ij}\, f\big(\ldots,h_{IJ}^{ij} g_I^i,\ldots , h_{IJ}^{ij}g_J^j, \ldots\big) .
\end{gather}
Since we are thinking of $g_I^i$ and $g_J^j$ as parallel transports of a gauge connection along edges emanating from the vertex to the endpoint,
it is natural to interpret such a dependence as the gluing of the edge $I$ incident on the vertex $i$ with the edge $J$ incident on the vertex $j$, with $(g_I^i)^{-1} g_J^j$ being the parallel transport from the vertex~$j$ to the vertex $i$ along a connecting edge.

In this way, we can attach to our states a graph label, based on the pattern of convolutions of the arguments, and identify GFT Fock states with a direct relation with the closed spin network wavefunctions in LQG. The precise connection with the (kinematical) LQG states is a delicate issue, and we refer to~\cite{Daniele2ndq} for a more complete discussion. First, the inner product that we are using here is the Fock space inner product, so that the graph label does not automatically identify orthogonal subspaces of the Fock space: states associated to dif\/ferent graphs, when they have the same number of quanta, are not necessarily orthogonal. Second, wavefunctions are completely symmetrised with respect to the permutations of vertices, while in the case of LQG the wavefunctions do not necessarily have to have special symmetrisation properties. Finally, GFT Fock states are not organised in terms of embedded graphs. The graphs we can associate to them are purely combinatorial objects, and make no reference to any background manifold structure, except to the group manifolds from which the data decorating the graphs are picked. Thus, they are closer to the setting of algebraic quantum gravity~\cite{AQG1,AQG2}, an approach for def\/ining canonical LQG without using embeddings or a dif\/ferentiable manifold structure.

Keeping in mind these subtleties, the Fock space language allows us to import many of the concepts of LQG into the GFT setting~\cite{Daniele2ndq}. In particular, one can immediately translate the operators that are normally def\/ined for spin networks into the second quantised language that we have introduced. In order to do so, we now consider the case $G = \SU$, the gauge group of the Ashtekar--Barbero formulation of classical general relativity~\cite{Ashtekar,Barbero} that is the basis of LQG.

The translation requires identifying the product of ladder operators that corresponds to the f\/irst quantised operators of interest. Hermitian operators are of particular signif\/icance, as they will store information about observables. In order to be Hermitian, these operators have to be built with (linear combinations of) strings of ladder operators possessing an equal number of creation and annihilation operators. According to the number of creation operators,
operators will be distinguished as one-body operators, two-body operators, and so on. The def\/inition of these operators is necessary to extract geometric information from states containing arbitrary superpositions of states associated to dif\/ferent graphs. The translation of intrinsically many-body operators (as, for instance, the holonomy computed on a loop incident on several vertices) is not immediate, as the nature of the GFT quanta as identical particles makes the identif\/ication of portions of the graphs a tricky issue, in absence of further degrees of freedom that can be used to distinguish between dif\/ferent kinds of vertices. Such degrees of freedom could, for example, correspond to matter f\/ields, or to an additional ``colour'' label of multiple GFT f\/ields (see Section~\ref{gencond} for some more discussion on colouring in GFT). For our purposes, one-body operators are enough to extract the relevant geometric information.

Generic one-body Hermitian operators have the form
\begin{gather*}
\widehat{X} = \int \dd g\, \dd g'\, X(g,g') \hphid(g) \hphi(g'), \qquad X(g,g') = \overline{X(g',g)} .
\end{gather*}

The simplest one-body operator is the number operator $\widehat{N}$, with $X=1$,
\begin{gather*}
\widehat{N} = \int \dd g\,\hphid(g) \hphi(g) .
\end{gather*}
This operator extracts a specif\/ic and unambiguous graph property, the number of vertices, or the average number of vertices in the case of a superposition of graphs.

Other common one-body operators that are relevant in the GFT context are the vertex volume operator $\widehat{V}$ and the plaquette area $\widehat{A}_I$.
They can be obtained by simply setting as $X(g,g')$ the LQG matrix elements of the desired operators, e.g.,
\begin{gather*}
\widehat{V} = \int \dd g\, \dd g'\, V^{\rm LQG}(g,g') \hphid(g) \hphi(g'),
\end{gather*}
where $V^{\rm LQG}(g,g')\equiv\langle g_I|V^{\rm LQG}|g'_I\rangle$ is the matrix element of the volume operator between single-vertex spin networks in LQG, and (replacing the matrix elements by a dif\/ferential operator)
\begin{gather}\label{areaop}
\widehat{A}_I = \int \dd g \, \hphid(g) \left(\sum_{m,n=1}^{3}\delta_{mn} \mathrm{E}^{m}_I\mathrm{E}^{n}_I\right)^{1/2} \rhd \hphi(g) .
\end{gather}
Here $\mathrm{E}^{n}_I$ are the LQG f\/lux operators, proportional to the Lie derivatives on $\SU$,
\begin{gather*}
\mathrm{E}^{n}_I f(g_J)= \lim_{\epsilon \rightarrow 0 } \iu\kappa\, \frac{d}{d\epsilon} f\big(e^{-\iu \sigma^n \epsilon/2}g_I\big),
\end{gather*}
where $\sigma^n$ is a Pauli matrix and the operator only acts on the $I$-th argument of the f\/ield, and~$\kappa$ is a parameter with dimensions of area, the ``Planck area'' of the theory, which in canonical LQG is usually set to $\kappa=8\pi\gamma\hbar G_{\rm N}$ where $\gamma$ is the Barbero--Immirzi parameter and $G_{\rm N}$ is the low-energy Newton's constant\footnote{Introducing $G_{\rm N}$ in this way, at the kinematical level in the microscopic theory, raises the question of whether~$G_{\rm N}$ can undergo renormalisation in LQG. The results in the later sections of this review will show how, in GFT condensate cosmology, the strength of the gravitational coupling to matter emerges from the coupling constants of the GFT action, so that~$G_{\rm N}$ has a rather dif\/ferent origin and is subject to renormalisation.}. Noting that contracting the Lie derivatives with the Killing form leads to the Casimir operator on $\SU$, (\ref{areaop}) can be simplif\/ied to
\begin{gather*}
\widehat{A}_I = \kappa\int \dd g\, \hphid(g) \sqrt{-\Delta_{g_I}} \hphi(g),
\end{gather*}
where $\Delta_{g_I}$ is the Laplace--Beltrami operator on $\SU$ acting on the $I$-th argument of $\hphi$. This shows explicitly how the (kinematical) area spectrum in LQG is related to the discrete spectrum of the Laplacian on $\SU$, multiplied by the parameter $\kappa$.

A basic property of one-body operators is that they are extensive: as it is immediate to realise using Wick's theorem, they generate a contribution from each of the quanta of the state on which they act. This aspect will play a role in our discussion of the condensate states and their geometrical content; by only using one-body operators in extracting geometric data from the states, we focus on quantities that are local at the level of each vertex or dual tetrahedron.

\subsection{Noncommutative dual formulation of GFT}\label{noncommSection}

Group f\/ield theories can be formulated in dif\/ferent equivalent representations, using suitably def\/ined Fourier transforms. The simplest one is the spin representation (using the termi\-no\-lo\-gy for~$\SU$), which is based on a Peter--Weyl decomposition. A slightly more sophisticated representation of the theory is the one that makes uses of noncommutative f\/luxes (see, e.g.,~\mbox{\cite{BaratinFlux,Guedes}}). The basic idea is to use a noncommutative Fourier transform that unitarily maps square-summable functions on the group to square-summable functions on the corresponding Lie algebra~$\mathfrak{g}$,{\samepage
\begin{gather*}
\mathcal{F}(f)(B) \equiv \int_G \dd g\, \ee_g(B) f(g),
\end{gather*}
where $\ee_g(B)$ are plane waves associated to the particular quantisation map chosen \cite{Guedes}.}

This allows a def\/inition of Fourier transformed f\/ield operators acting on the same Fock space,
\begin{gather*}
\hat{\tilde\varphi}(B_I) = \int \dd g \prod_{I=1}^4 \ee_{g_I}(B_I)\, \hphi(g_I),
\end{gather*}
where, corresponding to the four group elements $g_I$, $B_I = (B_1,\ldots ,B_4)$ with $B_I \in \mathfrak{g}$.

The peculiarity of the Lie algebra representation is that, instead of the ordinary product, a~noncommutative star-product has to be used in order to encode the noncommutative group multiplication. This product is def\/ined in terms of its action on plane waves,
\begin{gather*}
\ee_{g}(B) \star_B \ee_{h}(B) = \ee_{gh}(B).
\end{gather*}

While the Lie algebra representation presents a number of technical dif\/f\/iculties due to the nontrivial aspects of the star-product, it of\/fers a dual, and perhaps clearer geometric interpretation of the GFT quanta. Let us specialise again the discussion to $G=\SU$, where the Lie algebra variables $B_I$ can be naturally identif\/ied with the densitised triads integrated over the plaquettes dual to the edges attached to a GFT vertex, see~(\ref{bivector})\footnote{The construction of a noncommutative Fourier transform, used for the specif\/ic case of $\SU$ (and hence $\Spin$), can be extended, at least formally, to the case of weakly exponential Lie groups~\cite{Guedes}, including~$\SL$. What is so far missing is the explicit construction of the plane waves and the detailed analysis of their properties and ambiguities (related to the choice of a quantisation map) in their def\/inition \cite{Guedes}. Hence, so far there is no explicit construction of Lorentzian models based on f\/lux variables. However, there is also no obvious reason why such a construction should not be possible.}.

This interpretation is corroborated by the following observation. Using the (noncommutative) Dirac delta distribution in the Lie algebra representation, given by
\begin{gather*}
\delta_{\star}(B) = \int_G \dd g \,\ee_g(B),
\end{gather*}
we can easily see that the right gauge invariance of the group representation translates into a~closure condition of the f\/luxes, as
\begin{align}
\hat{\tilde\varphi}(B_I) &=  \int \dd g \int_G \dd h \prod_{I=1}^{4} \ee_{g_I}(B_I) \hphi\big(g_Ih^{-1}\big) = \int \dd g \int_G \dd h
\prod_{I=1}^{4} \ee_{g_Ih}(B_I) \hphi(g_I)\nonumber\\
&=  \hat{\tilde\varphi}(B_I) \star_{B_I} \delta_{\star}\left(\sum_{I=1}^4 B_I\right). \label{closure}
\end{align}
This closure condition $\sum_I B_I=0$ suggests that the state
\begin{gather*}
\ket{B_I} = \hat{\tilde\varphi}^\dagger(B_I) \ket{0}
\end{gather*}
can be interpreted as a closed tetrahedron, with $B_I$ being the bivector of the face~$I$.

The noncommutative counterpart of the convolution of group elements~\eqref{convolution} that leads to edges connecting dif\/ferent vertices is the insertion of a noncommutative Dirac delta, $\delta_\star(B_I^i + B_J^j)$, as can be seen by Fourier transforming~\eqref{convolution}. The insertion of a star-Delta function now can be interpreted as imposing the condition that the tetrahedron $i$ and the tetrahedron $j$ are glued along the faces $I$ and $J$, respectively. The fact that they are identif\/ied requires that the corresponding geometric data are properly matched, thus the condition $B_I^i = - B_J^j$.

In this way, states built with the noncommutative dual representation are naturally associated to three-dimensional simplicial complexes decorated with data that are directly related to the intrinsic Riemannian geometry. Therefore, they can be seen as triangulations of the spatial slices used in the computation of quantum gravity amplitudes. It should be kept in mind, however, that what we are gluing together are quantum tetrahedra. This will be evident whenever we will try to compute the action of geometric operators on these states. Despite the subtleties related to the noncommutative nature of the Lie algebra variables, the f\/lux representation allows a more transparent construction of states having specif\/ic continuum metric properties, for example corresponding to a particular spatially homogeneous metric as relevant in cosmology.

\subsection{Equations of motion in second quantisation}

The tools that we have introduced can be used to design interesting states, which can be interpreted in terms of spatial three-dimensional geometries. The last ingredient that we need is the one that will generate the sum over four-dimensional discrete geometries. This is achieved again in terms of the perturbative expansion of a suitable dynamical f\/ield equation. As a simple option, we can use the f\/ield equation from a classical action used in the partition function,
\begin{gather}
\frac{\delta S[\hphi,\hphid]}{\delta \hat\varphi(g)} \ket{\Phi} = 0 ,\label{EOM}
\end{gather}
but other choices are possible. What is important is that the f\/ield equation selecting the physical state $\ket{\Phi}$ can be split into a free part, which admits for instance the Fock vacuum as a solution, and an interaction term with the correct combinatorics.

\begin{figure}[t]\centering
\includegraphics[width=5cm]{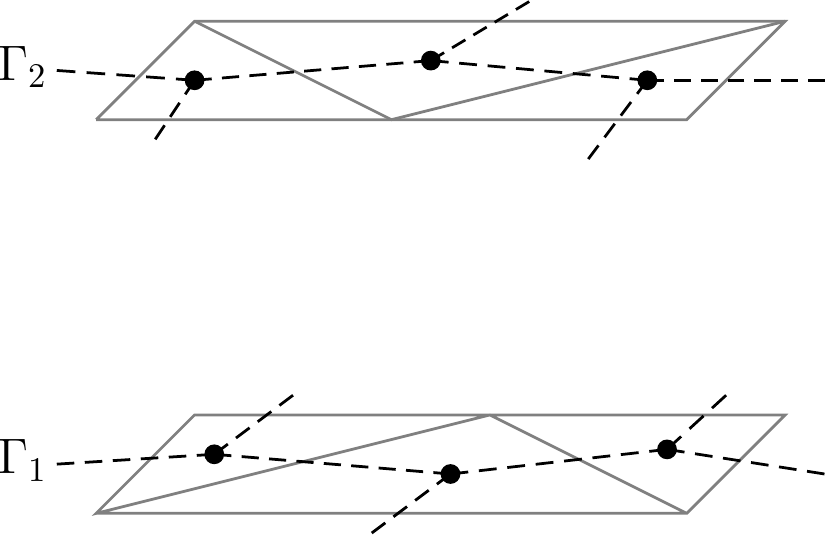} \hspace{2cm} \includegraphics[width=5cm]{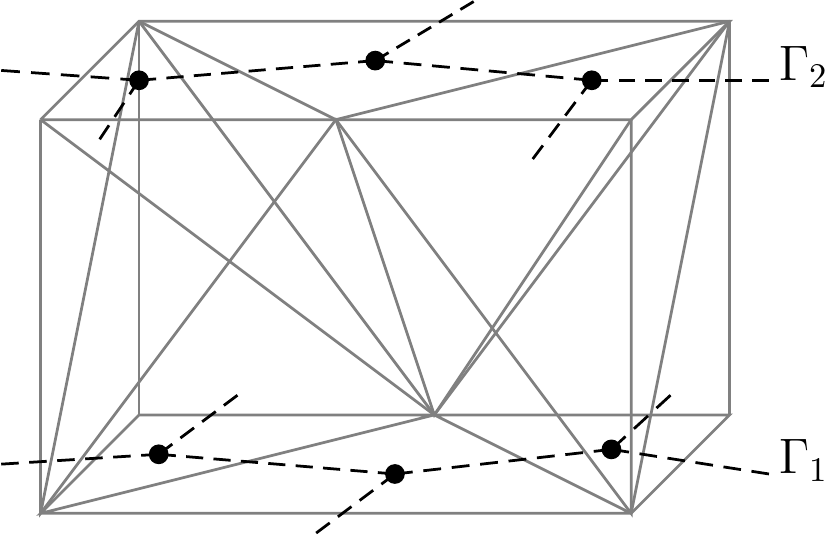}

\caption{Given two boundary states, the perturbative expansion of the correlation function \eqref{rigging} will generate all the simplicial complexes compatible with them, and the corresponding amplitude.} \label{perturbative}
\end{figure}

The perturbative expansion of the partition function is now translated into the expansion of~$\ket{\Phi}$ in powers of a coupling constant controlling the interaction term. Following the procedures of standard time-independent perturbation theory in quantum mechanics, one then recovers a~diagrammatic expansion of the various correlation functions in the f\/ield theory. To be more precise, in the previous sections we have seen how convolutions of f\/ield operators can be interpreted in terms of discrete geometric quanta, glued together to form specif\/ic geometrical structures. Consider then two sets of strings of creation operators, each associated to a graph, $\Gamma_1$, $\Gamma_2$, with data encoded in wavefunctions $f_1$, $f_2$. Let us denote these operators as $\widehat{\Psi}(\Gamma_i,f_i)$,
\begin{gather*}
\widehat{\Psi}(\Gamma_i,f_i) = \int (\dd g)^{\#V(\Gamma_i)} f_{i}(g^i, \Gamma_i) \prod_{j=1}^{\#V(\Gamma_i)} \hphid(g^j).
\end{gather*}
It is easy to see that a correlation function such as
\begin{gather}\label{rigging}
G(\Gamma_1,f_1; \Gamma_2, f_2)= \frac{\bra{\Phi} \widehat{\Psi}^\dagger(\Gamma_1,f_1) \widehat{\Psi}(\Gamma_2,f_2) \ket{\Phi}}{\braket{\Phi}{\Phi}},
\end{gather}
once the physical vacuum is treated perturbatively $\ket{\Phi} = \ket{0} + \sum\limits_{i=1}^{\infty} \lambda^i \ket{\Phi^{(i)}} $, can be rewritten in terms of a diagrammatic expansion, that can be interpreted in terms of four-dimensional discrete geometries having as boundaries the three-dimensional geometries dual to the graphs $\Gamma_1$, $\Gamma_2$, with the proper inclusion of bulk and boundary geometric data, as depicted schematically in Fig.~\ref{perturbative} for the case of a lower-dimensional theory.

The choice of the ingredients appearing in \eqref{EOM} (ladder operators and kernels) will then specify which graphs are included in the perturbative expansions of the correlation functions such as~\eqref{rigging}, together with the corresponding amplitudes. Gravitational models can be designed so that~\eqref{EOM} leads to graphs and amplitudes that match the ones of the spin foam expansion.

The correlation functions in \eqref{rigging} should not directly be interpreted in terms of transition amplitudes, as there is no notion of time evolution in this timeless framework. They should be viewed instead as the analogue of the matrix elements of the rigging map of LQG, enforcing the constraints that map kinematical states into physical states.

As in the functional formulation, a proper def\/inition of $\ket{\Phi}$ must pass through the examination of the perturbative series and its renormalisability, with all the related problems. The true advantage of an operatorial formulation is the possibility of developing a whole new approximation scheme for $\ket{\Phi}$ that is not based on a perturbation expansion, and this is the reason why we prefer this formulation of the theory.

If we view the theory def\/ined in \eqref{EOM} as a peculiar many-body problem of the kind encountered in condensed matter physics, we can try to abandon the perturbative expansion of the physical state and introduce a completely dif\/ferent approximation, based on a nonperturbative trial state $\ket{\Phi_{\mathrm{trial}}}$ that will include our guesses about the nature of the physical state of the theory, which will be presumably very dif\/ferent from the Fock vacuum that is used for the perturbative expansion. This is the point of view adopted when we are dealing with the GFT condensates, as it will be explained. It will be also manifest how this approach will make the investigation of the regime in which a large number of quanta are involved much more productive.

\subsection{Relation of GFT to tensor models and spin foam models}\label{sfrelation}

We have already mentioned that GFT is closely related to other approaches to the statistical mechanics of random geometries.

\looseness=1 The case of matrix models for random surfaces~\cite{diFrancesco} is perhaps the case that is understood with the greatest level of detail. The partition function, for a f\/inite size matrix model, is in fact an ordinary integral, and there are many dif\/ferent techniques that can be used to determine its behaviour. The critical properties of the model, describing the continuum limit of these models for discrete random surfaces, can be extracted from a properly def\/ined scaling limit, in which the coupling constants are tuned to critical values while the matrix size is sent to inf\/inity.

Tensor models (see \cite{GurauRyan,tensorreview} for reviews) are higher-dimensional generalisations of matrix models. They involve tensorial objects with higher rank, whose combinatorics is rich enough to permit the construction of Feynman diagrams that can be put in correspondence with simplicial complexes. As in the case of matrix models, tensor models are dealing only with the combinatorial structure of the Feynman diagrams. In recent years, a variety of results have been gathered about the combinatorial aspects of the partition function of tensor models, leading to a f\/irst understanding of their critical behaviour~\cite{BonzomDoubleScaling,GurauN}.

There is a strong relationship between tensor models and GFT. In a certain sense, one could view GFTs as tensor models with indices running on a~continuous set, a~Lie group, instead of a~discrete set. Conversely, a tensor model might be pictured as a GFT based on a discrete group as $\mathbb{Z}_N$, for instance. More generally, for compact groups, one can rewrite, via the Peter--Weyl decomposition, a GFT as a generalised tensor model involving various tensorial degrees of freedom, labelled by the irreducible unitary representations of the given group (with a cutof\/f to obtain a f\/inite index range). This reduces to a~conventional tensor model for $G={\rm U}(1)$.

GFT makes use of the same principles of the combinatorics of tensor models. In addition, however, the additional group-theoretic data give rise to a much richer variety of models, at f\/ixed combinatorics, corresponding to the possible ways to use this data to build the amplitudes. Most signif\/icantly, the additional data of GFT allow us to construct models which can be rewritten as simplicial path integrals for f\/irst-order gravitational theories \cite{BaratinOritiNC,BaratinOritiHP}, in particular Plebanski-type reformulations of general relativity that are adapted to the Ashtekar variables for LQG.

The analysis of the GFT partition function then has to go beyond the combinatorial aspects, and presents additional dif\/f\/iculties with respect to the partition function for matrix and tensor models. Indeed, the GFT formalism provides additional data in the amplitudes of the graphs, which are no longer characterised by the combinatorial structure alone. Therefore, the considerations leading to the identif\/ication of the classes of graphs responsible for the dominant contributions to the partition functions need to be revised. The hope is that the number of results obtained in the case of tensor models, particularly on the problem of renormalisability, phase transitions and the continuum limit, might be ultimately adapted and extended to GFT. The renormalisation of GFT is an ongoing research programme (see Section~\ref{renormsec}).

\looseness=-1 While GFT actions can obviously be of very general shapes, the special class built on generalisations of the Ooguri model plays an important role for the construction of candidates for quantum gravity models. The reason for this is the precise relationship between GFT and spin foam models, a proposal for the def\/inition of a discretised sum-over-histories version of quantum gravity. In turn, spin foams are closely related to loop quantum gravity, as spin foam amplitudes are related to histories of spin networks. In four spacetime dimensions, the question of whether spin foams and canonical LQG can be seen as def\/ining equivalent theories is a subtle issue and a~focus of current research~\cite{LQGSFreview}. The GFT formalism gives a dif\/ferent perspective on this issue: the canonical and covariant descriptions of GFT simply correspond to the standard dichotomy of def\/ining a quantum f\/ield theory through a Fock space/operator formalism or through the path integral. The two viewpoints on GFT can then be respectively be related to canonical LQG and spin foam models, without the need to directly map between canonical LQG and spin foams.

It is known that classical Einstein gravity without a cosmological constant can be formulated in a variety of ways using dif\/ferent sets of degrees of freedom and dif\/ferent variational principles (see, e.g.,~\cite{Peldan}). The Plebanski formulation starts from a topological BF theory and supplements it with a term imposing a certain set of constraints. These constraints make the theory equivalent to a Palatini (i.e., a~f\/irst order) formulation of vacuum general relativity. Indeed, the role of these constraints is to break the gauge symmetries of the topological theory, which would not possess the local degrees of freedom required by general relativity\footnote{It is worth mentioning that the three-dimensional case does not require the introduction of constraints, as three-dimensional general relativity without a cosmological constant can be written as a topological BF theory.}.

The spin foam approach to quantum gravity takes advantage of these results. The basic idea is to start from the quantisation of BF theories, suitably discretised on a cellular complex, and then to impose the necessary constraints that would turn a topological theory into a gravitational theory~\cite{baez}. In the topological theory, the data associated to the complexes (typically expressed in terms of representation labels) are allowed to take values in the whole range allowed by the gauge group involved, $\Spin$ or $\SL$. The imposition of the simplicity constraints restricts the allowed data that can be attached to a spin foam. In particular, when looking at the Lie algebra data, they are equivalent to the requirement that these data can be rewritten in terms of discretised tetrad f\/ields. The various proposals in the literature try to achieve this goal using a variety of related but dif\/ferent ideas and techniques~\cite{SFrev}.

There is a precise correspondence between spin foam models and GFT. For any given spin foam model, one can f\/ind a corresponding GFT action in a precise way~\cite{reiserov}: given the amplitude assigned to any given cellular complex by the spin foam model, there is a GFT action whose Feynman graphs, dual to cellular complexes, have matching amplitudes. Seen from the opposite direction, a GFT action def\/ines a spin foam model, once the partition function is rewritten in terms of the Feynman expansion,
\begin{gather}\label{expansion}
Z = \int\mathcal{D}\varphi\,\mathcal{D}\bar\varphi\,e^{- S[\varphi,\bar\varphi]}= \sum_\sigma \frac{\prod_i(\lambda_i)^{N_i(\sigma)}}{{\rm Aut}(\sigma)}  \mathcal{A}_\sigma.
\end{gather}
Here we are labelling the Feynman graphs with the corresponding dual cellular complexes $\sigma$, in order to make the correspondence with spin foam approach more transparent. We are also including the possibility that several dif\/ferent interactions terms are introduced, each with dif\/ferent kernels and dif\/ferent combinatorial structure, and each controlled by a dif\/ferent coupling constant $\lambda_i$. $N_i(\sigma)$ will be then the total number of vertices of type $i$ appearing in the Feynman diagram. ${\rm Aut}(\sigma)$ is the order of the automorphisms of $\sigma$. This is the combinatorial part of the amplitude. It can be shown by direct calculations that the amplitude $\mathcal{A}_\sigma$, the Feynman amplitude that the GFT model assigns to $\sigma$, can be represented as a spin foam model.

An important ingredient that is explicitly added by the GFT approach is a def\/inite prescription for generating weights for amplitudes for dif\/ferent cellular complexes in the sum over $\sigma$. These depend on the coupling constants of the GFT action, as (\ref{expansion}) shows explicitly.

We now present a concrete implementation of the ideas sketched so far, restricting the analysis to the case of models and data relevant for four-dimensional gravity. The starting point is the Ooguri model for $\Spin$ or $\SL$, which will be modif\/ied to include the restriction on the geometric data only. There are several similar but inequivalent constructions, as the imposition of simplicity constraints is not uniquely specif\/ied by the classical theory. Indeed, the translation of the classical, continuum constraints into the discrete setting faces various discretisation and ordering ambiguities related to the non-Abelian nature of the variables.  Furthermore, the simplicity constraints are characterised by a free parameter, the Barbero--Immirzi parameter.

The idea of GFT and spin foam models is to work with graphs decorated with the data of $\SL$ or $\Spin$ BF theories, and then to restrict the sums to only those data that correspond to the geometric degrees of freedom. For a f\/inite value of the Barbero--Immirzi parameter, one way to proceed is as follows. It is convenient to work with a f\/ield theory based on $\SU$, so that we can work directly with the geometric data and use the tools that have been built to deal with LQG spin networks. Following~\cite{DupuisLivine}, we can def\/ine a map $S$ from the space of functions of~$\SU$ into the space of functions of~$\Spin$ (or $\SL$)\footnote{Although, for the cases of interest, in the $\SL$ case they would be distributions.}. Denoting with $G_I$ the elements of the four-dimensional gauge group, the $\SU$ f\/ield is mapped into
\begin{gather*}
(S\varphi)(G_I) = \int \dd g\, S(G_I,g_J) \varphi(g_J) .
\end{gather*}
This map $S$ takes care of restricting the $\Spin$ data for the Feynman diagrams to geometric data only, and it is the part of the construction where the implementation of the simplicity constraints, encoded in the so-called fusion coef\/f\/icients~\cite{SFrev}, is located. The familiar expressions for the various spin foam models can be recovered by expanding these kernels in the appropriate basis functions (Wigner representation matrices, coherent states, noncommutative plane waves\footnote{As discussed in \cite{BaratinOritiNC,BaratinOritiHP}, it is possible to build GFT actions exploiting the relationship between their f\/lux representation, based on the noncommutative Fourier transform discussed in Section~\ref{noncommSection}, and simplicial path integrals that make direct use of f\/lux variables.}), but the explicit form of this kernel is not very illuminating.

The interaction term for the Ooguri model for $\Spin$, built with~\eqref{Ooguri}, can be used to def\/ine the interaction term for the $\SU$ theory,
\begin{gather*}
\mathcal{V}_{\text{SF}}[\varphi,\overline{\varphi}] = \mathcal{V}_{\text{Ooguri}}[S\varphi,\overline{S\varphi}] .
\end{gather*}
The presence of the maps $S$ will then guarantee that, in the Feynman diagrams, only the $\Spin$ data solving the simplicity constraints will be summed over. The Lorentzian case proceeds along similar lines, although some additional care has to be used as the various expressions might require regularisation, before
they can be manipulated.

While this is not the only possible way to proceed, it is the simplest one that allows us to work directly with geometric data and with the familiar~$\SU$ spin networks. This choice will also make it easier to keep track of the dif\/ferences between dif\/ferent prescriptions for the simplicity constraints. For the Barrett--Crane model, which corresponds to inf\/inite Immirzi parameter, the construction is slightly dif\/ferent as $\SU$ has to be replaced with the homogeneous space $\SL/\SU$ (or $\Spin/\SU\simeq\SO/{\rm SO}(3)\simeq\SU$ in the Riemannian case)~\cite{DePietri}.

\section{GFT condensates: kinematics}\label{kinesec}

After this discussion of the general formalism of GFT in terms of kinematics, dynamics, and the relation to LQG and spin foams, in this section we review the detailed construction and interpretation of condensate states in GFT. There are close similarities to how one proceeds in the case of a non-relativistic quantum f\/ield theory describing Bose--Einstein condensates \cite{BECbook}, but also important dif\/ferences due to the fact that we are dealing with a f\/ield theory of quantum geometry as a (background-independent) proposal for quantum gravity. We largely follow the constructions and discussion given in the original papers \cite{PRL,JHEP}, on which most of the work so far has been based. In Section~\ref{gencond} we will show how these constructions can be generalised and extended to states that might be closer to those of physical interest in quantum gravity, and that have the advantage of also describing situations of spherical symmetry.

\subsection{Def\/initions and simplest condensates}\label{simplecondsec}

A condensate phase of a quantum f\/ield theory can be characterised by an {\em order parameter}, associated to a non-zero expectation value of a f\/ield operator in this phase which would be zero in the standard Fock vacuum. In the simplest case, this order parameter can be associated with the ``condensate wavefunction'', i.e., a~complex-valued function on the conf\/iguration space of the basic quantum f\/ield that fully determines the quantum state.

More concretely, we can focus on a particular subset of the space of general $N$-particle states
\begin{gather*}
|\Psi\rangle:=\int (\dd g)^N \Psi\big(g_I^1,\ldots,g_I^N\big) \bigotimes_{i=1}^N|g_I^i\rangle
\end{gather*}
in which the wavefunction $\Psi$ decomposes into a product of functions depending on fewer arguments. Physically, this means neglecting correlations between dif\/ferent particles either altogether or beyond a certain order (e.g., beyond 2-particle or 3-particle correlations). The two simplest cases are of particular interest: in the f\/irst, $\Psi$ decomposes into one-particle wavefunctions,
\begin{gather}\label{productstate}
\Psi\big(g_I^1,\ldots,g_I^N\big)=\prod_{i=1}^N \psi\big(g_I^i\big),
\end{gather}
whereas in the second it involves two-particle correlations (this is known as the {\em Bogoliubov approximation} in the context of Bose--Einstein condensates~\cite{BECbook}),
\begin{gather}\label{2particlecorr}
\Psi\big(g_I^1,\ldots,g_I^N\big)=\prod_{i=1}^{N/2} \psi\big(g_I^{2i-1},g_I^{2i}\big) + {\rm permutations}
\end{gather}
where $N$ is even and the permutations are needed to ensure that the wavefunction $\Psi$ is symmetric under the exchange of any two vertex indices, compatible with the bosonic statistics for the GFT. This symmetry will be automatically implemented once bosonic f\/ield operators are used.

States of the type (\ref{productstate}) or (\ref{2particlecorr}) could be used as simple approximations to a condensate, and capture much of the relevant physics of condensates. However, they do not satisfy all of the desired properties; in particular, given that they describe states of f\/ixed particle number, the expectation value of the f\/ield operators $\hat\varphi$ and $\hat\varphi^{\dagger}$ vanish. We hence need to use states that are superpositions of states of the type (\ref{productstate}) or (\ref{2particlecorr}) for dif\/ferent $N$, with appropriate weights in the sum, and are led to the def\/inition of a {\em single-particle condensate} state
\begin{gather}\label{singlecon}
|\sigma\rangle:=\mathcal{N}(\sigma)\exp\left(\int \dd g\, \sigma(g_I) \hat\varphi^\dagger(g_I)\right)|0\rangle,
\end{gather}
where $\mathcal{N}(\sigma)=\exp(-\half\int \dd g\, |\sigma(g_I)|^2)$ is a normalisation factor, and a {\em dipole condensate} state
\begin{gather}\label{dipolecon}
|\xi\rangle:=\mathcal{N}(\xi)\exp\left(\int \dd g\,\dd h\, \xi\big(g_I^{-1} h_I\big) \hat\varphi^\dagger(g_I) \hat\varphi^\dagger(h_I)\right)|0\rangle,
\end{gather}
with $\mathcal{N}(\xi)$ again a normalisation factor (given explicitly in \cite{JHEP}). Note the def\/inition of $\xi$ in~(\ref{dipolecon}), instead of an in principle possible more general wavefunction $\psi(g_I,h_I)$. Because $\hat\varphi^\dagger(g_I)= \hat\varphi^\dagger(g_Ih)$, the functions $\sigma$ and $\xi$ satisfy $\sigma(g_I)=\sigma(g_Ih)$ and $\xi(g_I)=\xi(kg_Ih)$ for all $h,k\in G$. For~(\ref{singlecon}) to also represent a state invariant under a ``gauge transformation'' acting simultaneously on the open ends of the spin network represented by an open vertex, we also impose $\sigma(g_I)=\sigma(kg_I)$ for all $k\in G$. (\ref{dipolecon}) is built out of elementary ``dipoles'' that correspond to gauge-invariant states in LQG, and are characterised by a single wavefunction $\xi(g_I)$ on $G \backslash G^4 /G$, as would be the possible states on a ``dipole'' in usual LQG~\cite{SFlectures}.

The states (\ref{singlecon}) and (\ref{dipolecon}) satisfy the property of {\em wavefunction homogeneity} which requires the state to be determined by a single-particle wavefunction ($\sigma$ or~$\xi$). They are superpositions of states of the type (\ref{productstate}) and (\ref{2particlecorr}) for dif\/ferent particle numbers, and hence rather dif\/ferent from the states usually considered in LQG which are typically built on a f\/ixed graph.

An important property of the state (\ref{singlecon}) is that it is an eigenstate of the GFT f\/ield operator,
\begin{gather*}
\hat\varphi(g_I)|\sigma\rangle = \sigma(g_I)|\sigma\rangle.
\end{gather*}
From this it immediately follows that $\langle\hat\varphi(g_I)\rangle=\sigma(g_I)$, so that $\sigma(g_I)$ represents the order parameter, and that more generally, any normal-ordered $n$-point function involving $\hat\varphi(g_I)$ and $\hat\varphi^\dagger(g_I)$ can be evaluated by replacing $\hat\varphi$ by $\sigma$ and $\hat\varphi^\dagger$ and $\bar\sigma$. In this very simplest approximation, there are no correlations between the GFT quanta, and in the mean-f\/ield approximation when f\/luctuations can be ignored, $\sigma(g_I)$ directly represents a classical GFT f\/ield conf\/iguration. This property will be important in Section~\ref{interpretsec} when the geometric interpretation of~(\ref{singlecon}) will be discussed.

As is apparent from the appearance of an explicit normalisation factor in (\ref{singlecon}) and (\ref{dipolecon}), these states are normalisable for appropriate choices of the functions $\sigma$ and $\xi$, and hence elements of the GFT Fock space def\/ined in Section~\ref{gftsec}. For (\ref{singlecon}), normalisability of the state is equivalent to a~f\/inite expectation value of the number operator,
\begin{gather}\label{particlen}
N:=\langle\sigma|\widehat{N}|\sigma\rangle = \langle\sigma| \int \dd g\, \hat\varphi^\dagger(g_I)\hat\varphi(g_I)|\sigma\rangle = \int \dd g \, |\sigma(g_I)|^2 < \infty.
\end{gather}
The thermodynamic limit $N\rightarrow\infty$ is described by states that are no longer in the original GFT Fock space: this is standard in quantum f\/ield theory, where in the limit corresponding to a~phase transition one needs to change representation to a dif\/ferent, unitarily inequivalent Hilbert space. Here $N\rightarrow\infty$ corresponds to a continuum limit in which such a transition is reached.

In practice, the dif\/ference between f\/ixed particle number as in~(\ref{productstate}) and f\/luctuating particle number as in (\ref{singlecon}) becomes unimportant for large $N$: an elementary calculation shows that
\begin{gather*}
\left(\Delta N\right)_{|\sigma\rangle}:= \sqrt{\langle\sigma|\widehat{N}^2|\sigma\rangle-(\langle\sigma|\widehat{N}|\sigma\rangle)^2}=\sqrt{\langle\sigma|\widehat{N}|\sigma\rangle}=\sqrt{N}
\end{gather*}
and hence the relative uncertainty $(\Delta N)/N\sim 1/\sqrt{N}$ vanishes in the thermodynamic limit; this is the well-known equivalence of the canonical and grandcanonical ensembles in this limit.

The dipole condensate state (\ref{dipolecon}) is no longer a coherent state for the GFT f\/ield operator, but closer to squeezed states that appear in quantum optics \cite{opticbook}. While a ref\/inement of (\ref{singlecon}) that incorporates the usual notion of gauge invariance in LQG and contains two-particle correlations, is already harder to do computations with; no simple closed expression for general $n$-point functions or quantities such as the average particle number is known. Approximate expressions for correlation functions are discussed in \cite{JHEP}, and can be used for estimates of the error that one makes in using them.

The def\/inition of the states (\ref{singlecon}) and (\ref{dipolecon}) is mainly motivated by their nice properties from the perspective of quantum f\/ield theory and by the analogy to the well-studied case of Bose--Einstein condensates. They assume a truncation to one- and two-particle correlations respectively, which can be seen as corresponding to the situation of cosmological relevance, where we focus on geometries that are spatially homogeneous or nearly homogeneous, so that spatial gradients can be ignored in a f\/irst approximation. However, the truncation also means that these states cannot be interpreted in the sense of an extended graph in LQG: they would correspond to ``graphs'' consisting of a~large number of disconnected components formed by a~single vertex or dipole, respectively. Any topological information normally also contained in the graph must come from elsewhere. One way to remedy this situation is to def\/ine generalised condensates that are associated to a graph of f\/ixed topology, as we will outline in Section~\ref{gencond}. Another is to make an assumption for the topology of space and require self-consistency with the dynamics: e.g., if the ef\/fective dynamics of isotropic and homogeneous geometries is given by the Friedmann equation with positive spatial curvature, one infers that space should have the topology of a 3-sphere. This viewpoint, advocated in~\cite{JHEP} and in some sense closer to the formalism of canonical LQG \cite{LQGbook1,LQGbook2}, will be outlined in Section~\ref{interpretsec}.

\subsection{``Condensate wavefunction'' vs.~``wavefunction of the universe''}\label{wavefsec}

The mean f\/ield $\sigma(g_I)$ appearing in the def\/inition (\ref{singlecon}) is often referred to as the ``condensate wavefunction''. Indeed, it is a~wavefunction in the sense that $|\sigma\rangle$ is a superposition of states with all possible particle numbers, each of which is def\/ined by a product wavefunction of the form~(\ref{productstate}) where $\psi$ is replaced by $\sigma$: $\sigma$ is the wavefunction of the ``ground state'' that the GFT quanta are supposed to be condensed into (again, in this mean-f\/ield approximation where f\/luctuations over this condensed phase are ignored).

In most other respects, it can be quite misleading to think of $\sigma$ as a usual quantum-mechanical wavefunction. First of all, the overall normalisation of $\sigma$ is not arbitrary as in quantum mechanics, but corresponds to the average particle number (see~(\ref{particlen})). Indeed it is often useful to adopt a parametrisation $\sigma(g_I)=\sqrt{N} \sigma_0(g_I)$ where $\sigma_0$ is normalised to one~\cite{BECbook}.

More importantly, there is no superposition principle and no direct probability interpretation associated with $\sigma(g_I)$, as the dynamics of~$\sigma$ is typically governed by nonlinear (and nonlocal) equations. At f\/irst sight, this may appear confusing: are we proposing a radical nonlinear, nonlocal modif\/ication of quantum mechanics for GFT? It should be clear, and will become clearer in Section~\ref{dynsec} below, that this is not the case, and all dynamical equations are linear equations on the GFT Fock space. However, the correspondence between a condensate state and its ``wavefunction'' is clearly nonlinear,
\begin{gather*}
|\sigma\rangle + |\sigma'\rangle \neq |\sigma+\sigma'\rangle,
\end{gather*}
and linear equations on the Hilbert space become nonlinear equations for $\sigma$. Whatever the precise form of the ef\/fective dynamics for~$\sigma$, it will not make sense to directly interpret them as a Schr\"odinger-type or Wheeler--DeWitt-type equation for a~``wavefunction of the universe'' as in standard quantum cosmology \cite{oldQCreview}, and common tools in quantum cosmology such as the WKB approximation will have to be reinterpreted according to the r\^{o}le of $\sigma$ as a mean-f\/ield conf\/iguration of the quantum f\/ield, rather than a single-particle (``single-universe'') wavefunction.

To make this more precise, let us compare with the situation in condensed matter physics. Here the ``condensate wavefunction'' describing the simplest approximation to the Bose--Einstein condensate can be written as \cite{BECbook}
\begin{gather}\label{BECPsi}
\Psi(\vec{x})=\sqrt{\rho(\vec{x})} \exp(-\im\theta(\vec{x})),
\end{gather}
where $\rho(\vec{x})$ def\/ines the density and $\theta(\vec{x})$ the velocity potential (for irrotational f\/luids, the velocity is $\vec{v}\propto\vec\nabla\theta$) of the superf\/luid formed by the condensate. $\Psi$ clearly has no probability interpretation; it already describes a classical limit, the mean-f\/ield approximation, in which it corresponds to a classical f\/ield conf\/iguration (given by two real functions $\rho,\theta$ or one complex function~$\Psi$).

\looseness=1 A WKB approximation applied to (\ref{BECPsi}) would correspond to a regime in which the f\/luid density varies slowly over space while the velocity $\vec{v}$ of the f\/luid is large, and such an appro\-xi\-mation will not be valid in general, if we remember that (\ref{BECPsi}) corresponds to a ground-state conf\/iguration. For instance, for the very simplest case of a non-interacting Bose gas, $\Psi(\vec{x})$ is simply constant over space (over a box in which the gas is conf\/ined) but the velocity $\vec{v}$ va\-nishes.

Phrased dif\/ferently, a WKB approximation would correspond to a regime in which the fundamental degrees of freedom, i.e., the individual atoms in the case of the condensate or the degrees of freedom of quantum geometry in the case of GFT, already behave semiclassically. While semiclassical properties of the quantum state are an important consistency criterion in quantum cosmology (see, e.g.,~\cite{jeanluc}), if quantum cosmology arises from the hydrodynamics of quantum gravity semiclassicality is not a restriction on the ``wavefunction'' $\sigma$ but rather a statement about the validity of an ansatz of the form (\ref{singlecon}), i.e., of the mean-f\/ield approximation. Early works on GFT condensates advocated the use of a WKB approximation \cite{PRL,JHEP}, but it was understood in \cite{SteffenCQG,SteffenPRD,NJP} that this is not really justif\/ied. In this review, we will focus on studies of the dynamics of GFT condensates that do not assume the validity of a WKB regime. Nevertheless, interesting results such as \cite{LQCfromGFT} where the dynamics of LQC and the ef\/fective dynamics of GFT condensates could be matched in a WKB regime show that the role of WKB solutions in this setting deserves to be better understood.

Because the ``condensate wavefunction'' is a mean-f\/ield conf\/iguration rather than a quantum-mechanical wavefunction, there is no obvious map between the formalism of GFT condensates and the Hilbert space formalism of Wheeler--DeWitt quantum cosmology or LQC, but such a map is not required; what is needed is a way to extract cosmological predictions from the ef\/fective dynamics of such condensates. For this it has proven advantageous to focus on expectation values of geometric operators that correspond to cosmological observables, as we will detail in Section~\ref{dynobssec} and below. Viewed more practically, the absence of a Hilbert space structure in this hydrodynamic setting might be seen as an advantage rather than a problem in quantum cosmology, where a general ``wavefunction of the universe'' has serious interpretational problems and one usually uses a class of special, highly semiclassical states, so that the full Hilbert space seems rather redundant when one is interested in the physical predictions of quantum cosmology. In our setting, the semiclassical properties expected from the universe are already encoded in the use of a restricted class of (condensate) states.

\subsection{Interpretation in terms of geometry}\label{interpretsec}

So far, the def\/inition of GFT condensate states has mainly been motivated by the analogy with Bose--Einstein condensation in condensed matter physics. In order to relate this setting to quantum cosmology, we also need to show how states such as (\ref{singlecon}) or (\ref{dipolecon}) can be interpreted in terms of spacetime geometry, i.e., in terms of (a probability distribution over) continuum metrics and/or Ashtekar--Barbero connections. For this, a slight detour is necessary in which we discuss in more detail the geometric interpretation of the conf\/iguration space of the GFT f\/ield $\hat\varphi(g_I)$ and, accordingly, the wavefunctions $\sigma$ and $\xi$ appearing in (\ref{singlecon}) and (\ref{dipolecon}). Any interpretation proposed in this context is motivated by the close relation of GFT to LQG and spin foam models, which do give a geometric interpretation to the group-theoretic data associated to the states. Eventually, it will have to be shown that such an interpretation is justif\/ied, i.e., that the ef\/fective continuum geometry def\/ined from it is governed by the dynamics of a gravitational theory (Einstein's equations, or a generalisation of them), couples correctly to matter, and satisf\/ies a notion of dif\/feomorphism invariance. We will outline how this could be done for GFT condensates in the discussion of dynamics in Section~\ref{dynsec}.

What we are looking for is an ``inverse map'': given the parallel transports $g_I$ of a connection along four edges meeting at a central vertex,
\begin{gather*}
g_I \sim \mathcal{P}\exp \left(\int_{e_I} A\right)
\end{gather*}
or, in the dual non-commutative formulation introduced in Section~\ref{noncommSection}, four Lie algebra ele\-ments~$B_I$ corresponding to bivectors associated to faces of a dual tetrahedron (where we assume that simplicity constraints are either trivial, for $G=\SU$, or have been imposed properly),
\begin{gather}\label{bivector}
B^{AB}_I \sim \int_{\triangle_I} e^A\wedge e^B,
\end{gather}
how do we reconstruct an (approximate) connection $A$ or a coframe f\/ield $e$ compatible with this data? This is a common problem in discrete approaches to quantum gravity, and hard to address \cite{bombe1,bombe2}. Our task is somewhat easier than in general because we are looking at specif\/ic states with high symmetry, and expect the approximate continuum description to correspond to a homogeneous or almost homogeneous geometry.

The following interpretation of the bivectors (\ref{bivector}) in terms of a continuum metric was given in \cite{JHEP}. First, the mean f\/ield $\sigma(g_I)$\footnote{Here and in the following, $\sigma(g_I)$ stands for a general ``condensate wavefunction'' with the required symmetries, and in particular can also stand for the function $\xi(g_I)$ used in~(\ref{dipolecon}).} has two independent symmetries $\sigma(g_I)=\sigma(hg_Ik)$ for all $h,k\in G$, which imply that its noncommutative Fourier transform,
\begin{gather}\label{ncsigma}
\tilde\sigma(B_I):=\int \dd g  \prod_{I=1}^4 e_{g_I}(B_I)\,\sigma(g_I),
\end{gather}
satisf\/ies (compare with the property (\ref{closure}) of the GFT f\/ield itself)
\begin{gather*}
\tilde\sigma(B_I)=\delta_{\star}\left(\sum_I B_I\right)\star\tilde\sigma(B_I),\qquad \tilde\sigma(B_I)=\tilde\sigma\big(kB_Ik^{-1}\big)\qquad\forall\, k\in G.
\end{gather*}
These symmetries correspond to two symmetries in the continuum: the f\/irst is the {\em closure constraint} implying that the four faces close to form a~geometric tetrahedron, and the second is the invariance under local frame rotations which are classical discrete gauge transformations.

The next step is to assume that the continuum f\/ields to be reconstructed vary slowly over the scale of the tetrahedron, as must be the case for the discretisation to approximately represent them. This means we can replace $B^{AB}_i = \frac{l^2}{2} {\epsilon_i}^{jk} e^A_j e^B_k$ in terms of three discrete triad vectors~$e^A_i$, where $i=1,2,3$ and~$l$ is a coordinate length associated to the triangles that we leave free for now; $B_4$ is replaced by $-\sum_i B_i$ using the closure constraint. The gauge-invariant conf\/iguration space for each tetrahedron, invariant under the transformations $B_i\mapsto kB_ik^{-1}$, is six-dimensional and can be parametrised by the six ``metric components''
\begin{gather}\label{metriccomp}
g_{ij}=e^A_i e^B_j\eta_{AB}=\frac{1}{4l^2{\rm tr}(B_1B_2B_3)}{\epsilon_i}^{kl}{\epsilon_j}^{mn}\tilde{B}_{km}\tilde{B}_{ln},\qquad \tilde{B}_{ij}:={\rm tr}(B_i B_j),
\end{gather}
where $\eta_{AB}$ is the appropriate invariant inner product. The quantities $g_{ij}$, constructed out of~(\ref{bivector}), should be scalars under dif\/feomorphisms. They represent the 3-metric in a given frame that transforms covariantly under dif\/feomorphisms, at a given f\/ixed point in the tetrahedron (which can be chosen for convenience to be one of the vertices). The continuum interpretation of~(\ref{metriccomp}) is f\/ixed after specifying this frame. In essence, the freedom in choosing this frame corresponds to the freedom in rotating tetrahedra in an embedding into a spatial manifold. This freedom is still there because our construction is purely local on the level of each individual tetrahedron and disregards any connectivity information in the states, as seems appropriate when considering the simple condensates of Section~\ref{simplecondsec}.

In order to def\/ine a notion of spatially homogeneous geometries, we now consider an embedding of a given conf\/iguration of tetrahedra into a continuum manifold of topology~$\mathfrak{G}/X$ where~$\mathfrak{G}$ is a~3-dimensional Lie group acting transitively on the manifold, and $X$ can be a~discrete subgroup~\cite{JHEP}. $\mathfrak{G}$ def\/ines the notion of homogeneity, and also suggests a natural choice of frame, given by a basis of left-invariant vector f\/ields. The $g_{ij}$ in~(\ref{metriccomp}) are then interpreted as metric coef\/f\/icients in this frame, so that a spatially homogeneous geometry is precisely one in which $g_{ij}$ are all constant in the continuum, or the same for a large number of tetrahedra in the discrete. This then justif\/ies the choice of state~(\ref{productstate}) or~(\ref{singlecon}) in order to describe continuum spatially homogeneous geometries, where the continuum is strictly obtained in the limit $N\rightarrow\infty$.

Our use of an embedding in this context is very much analogous to how one interprets spin network states in canonical LQG as encoding (distributional) continuum geometric data on a given, continuum spatial hypersurface~\cite{LQGbook1}. The topological information that one may associate to a~collection of~$N$ tetrahedra (i.e., a disjoint union of $N$ three-balls) plays no role in this continuum interpretation, and there would be no consistent way of using it if we are looking at superpositions of dif\/ferent~$N$, e.g.,~(\ref{singlecon}). It seems plausible that, if cosmology and continuum geometry emerge from a~hydrodynamic approximation to quantum gravity, the topology of space should also be emergent rather than determined by microscopic details, such a specif\/ic choice of graph. Indeed, the fact that one might think of a superf\/luid fundamentally as a collection of disconnected quantum atoms plays no role at the hydrodynamic level. This viewpoint leaves open the question of how to choose the embedding manifold, as we will discuss below. One might try to incorporate a choice of topology into the def\/inition of the state: in Section~\ref{gencond}, we will discuss a~dif\/ferent class of condensates with def\/inite topological interpretation, which may be seen as an alternative proposal.

In (\ref{metriccomp}), we have focused on the metric which is usually more important in cosmology, but a~similar construction of a continuum connection from the group elements $g_I$, taking into account the invariance under $g_I\mapsto hg_I k$, is possible. As discussed in some detail in \cite{SteffenCQG} for $G=\SU$, if GFT condensates def\/ine spatially homogeneous continuum geometries, the phase space of a~tetrahedron can be mapped to the phase space of spatially homogeneous continuum geometries, with the quotient space $\SU\backslash\SU^4/\SU\simeq\SU^3/{\rm Ad}_{\SU}$ of gauge-invariant group variables corresponding to continuum homogeneous connections and the quotient $\mathfrak{su}(2)^{\oplus 3}/{\rm Ad}_{\SU}$ of gauge-invariant bivector variables corresponding to continuum homogeneous metrics. This observation makes explicit in which sense a tetrahedron can be seen as a ``chunk of space'' with the same geometric data as a continuum geometry, and it provides an important foundation for the idea of cosmology as hydrodynamics of GFT: it shows that GFT condensate states, based on the idea of wavefunction homogeneity, capture exactly the degrees of freedom needed for homogeneous cosmology, i.e., minisuperspace geometric degrees of freedom. The ambiguities in interpreting these degrees of freedom are related to the ambiguity of def\/ining an isomorphism between GFT data and continuum cosmological variables.

Using (\ref{metriccomp}) gives a direct interpretation of the gauge-invariant geometric variables of a~tetrahedron in terms of a~continuum metric, which is relatively simple and local at the level of each tetrahedron. There are some technical dif\/f\/iculties in def\/ining an operator version of~(\ref{metriccomp}), f\/irst from def\/ining inverse operators, and also from operator ordering ambiguities coming from non-commutativity of the $\tilde{B}_{ij}$~\cite{SteffenCQG}. Barring these, one could now def\/ine an ef\/fective (spatially homogeneous) continuum metric out of expectation values of such operators.

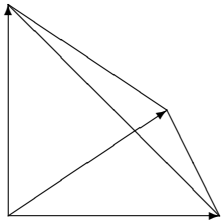
\begin{figure}[t]\centering
\begin{picture}(100,80)
\put(0,0){\vector(0,1){80}}\put(0,0){\vector(1,0){80}}\put(0,0){\vector(3,2){60}}
\put(0,80){\line(1,-1){80}}\put(60,40){\line(1,-2){20}}\put(60,40){\line(-3,2){60}}
\end{picture}
\caption{An ``isotropic tetrahedron'' with three pairwise orthogonal edges.}\label{tetrahed}
\end{figure}

There are however several issues with (\ref{metriccomp}), mainly related to its need to be supplemented by a choice of frame in which the $g_{ij}$ def\/ine the 3-metric. First of all, even though def\/ining this frame by three left-invariant vector f\/ields is rather natural, the notion of left-invariance itself relies on the choice of isometry group $\mathfrak{G}$, with the possible choices corresponding to the homogeneous cosmological models according to the Bianchi classif\/ication (see, e.g.,~\cite{bojobook}). The viewpoint taken in~\cite{JHEP} is that this choice should be f\/ixed by self-consistency: one does not know initially which of the Bianchi models is described by a particular condensate state, but its ef\/fective dynamics will pick out at most one, as the Bianchi models all dif\/fer by their ``anisotropy potential'' in the Hamiltonian constraint~\cite{bojobook}. While this indeed suggests an unambiguous association of condensate states with Bianchi universes, a more fundamental understanding why a particular Bianchi geometry would be selected is so far lacking. Perhaps more importantly, the def\/inition~(\ref{metriccomp}) uses only the bivectors $B_1$, $B_2$ and $B_3$, as $B_4$ has been eliminated using the closure constraint $\sum_I B_I=0$. The four faces of the tetrahedron are not treated ``democratically'': an isotropic tetrahedron would be one with three pairwise orthogonal edges of equal length, rather than a fully equilateral one, see Fig.~\ref{tetrahed}. One might prefer constructions in which isotropy means equilateral tetrahedra~\cite{VolumeDynamics}. The closure constraint also breaks the ${\rm O}(3)$ symmetry of continuum geometry under a~redef\/inition of the basis of left-invariant vector f\/ields; under such a~redef\/inition, $e_i\mapsto {O_i}^j e_j$ and $B_i\mapsto \pm {O_i}^j B_j$ which in general does not preserve $\sum_I B_I=0$. This suggests that there should be a~preferred basis of left-invariant vector f\/ields, and for Bianchi models such a basis exists: it is the basis (def\/ined up to the action of a~permutation group~$\mathfrak{S}_3$) in which the metric is diagonal, $g_{ij}=a^2_i\delta_{ij}$ with no summation over~$i$. Hence, for consistency, we are required to assume that of\/f-diagonal components of~(\ref{metriccomp}) vanish, at least as expectation values, which gives a restriction on possible condensate states to def\/ine Bianchi models. Again, this is not a problem in practice, but might suggest that the construction is not general enough\footnote{It also appears that a similar issue would apply to connection variables, leading to the conclusion that these would also have to be diagonalised. Then one would be restricted to class~A Bianchi models, for which a~simultaneous diagonalisation of the spatial metric and its conjugate momentum is possible and a corresponding Hamiltonian formulation exists~\cite{bojobook}.}.

So far, much of the work on GFT condensates has used geometric observables that do not rely on (\ref{metriccomp}) but are built out of more basic operators borrowed from LQG (see Section~\ref{2ndquantsec}). The perhaps most basic such operator is a volume operator, which in the second quantised formalism simply corresponds to the total volume contained in a state. The def\/inition of the volume operator in LQG is somewhat subtle, with dif\/ferent def\/initions arising from dif\/ferent regularisation schemes \cite{volume1,volume2} and possible issues with a consistent semiclassical limit \cite{cecthom}; these issues are less important for tetrahedra, for which good agreement with Bohr--Sommerfeld quantisation has been found \cite{eugeniohal}. Given a def\/inition of the volume operator, one can compute the expectation value or higher moments of the volume of the universe. For a homogeneous, isotropic geometry, this directly def\/ines a scale factor $a$, proportional to the third root of this total volume. For more general geometries, the total volume is just the most coarse-grained geometric observable. Hence, for an identif\/ication of the total volume with a quantity proportional to $a^3$ to make sense, one needs to be conf\/ident that the state does def\/ine a homogeneous and isotropic geometry. The embedding picture above gives an argument for why condensates def\/ine spatially homogeneous geometries, but the notion of isotropy must come from elsewhere if one does not want to use the def\/inition (\ref{metriccomp}). For instance, one may def\/ine isotropy by equilateral tetrahedra, see Section~\ref{LQCrelation} below.

As discussed in Section~\ref{2ndquantsec}, the GFT Fock space formalism also allows the def\/inition of ``total area operators'', one-body operators that are simpler to def\/ine than the total volume as they are directly built out of the basic f\/lux operators in LQG. They take the form
\begin{gather*}
\widehat{A}_I = \int \dd g\, \hat\varphi^\dagger(g_J) \,\sqrt{-\Delta_{g_I}}\hat\varphi(g_J)
\end{gather*}
and describe quantities such as ``the total area of all $I$-th faces''. Their geometric interpretation is not entirely clear in general, since one seems to be summing over areas of disconnected surfaces, which is dif\/ferent from LQG operators that correspond to the area of a 2-surface in space. For a homogeneous and isotropic geometry, however, again the expectation value $\langle \widehat{A}_I \rangle$ of such an operator has to be proportional to the square of the scale factor. It turns out that $\langle \widehat{A}_I \rangle$ divided by $N^{1/3}$ can be identif\/ied with a cosmological scale factor, so that
\begin{gather}\label{scalefdef}
a^2 V_0^{2/3} := \big\langle\widehat{A}_I\big\rangle N^{-1/3},
\end{gather}
where $V_0$ is a f\/iducial coordinate volume, akin to the volume of the ``f\/iducial cell'' appearing in homogeneous models for quantum cosmology, such as LQC \cite{LQCrev1,LQCrev2}. There are several arguments justifying the factor $N^{-1/3}$ in (\ref{scalefdef}). First, consistency requires that the Poisson brackets of the ef\/fective cosmological quantities reduce to the ones of classical general relativity in the limit of low curvature \cite{SteffenJHEP}, which f\/ixes this factor. One can also motivate it heuristically by thinking of a~large ``f\/iducial cell'' encompassing part of the universe one is interested in, itself composed of $N$ elementary cells. One would associate the cosmological area $a^2 V_0^{2/3}$ to one of the sides of the f\/iducial cell. The operator $\widehat{A}_I$, however, takes into account all elementary cells, and hence overcounts by a factor $N^{1/3}$, the number of ``layers'' within the f\/iducial cell. Similar ideas have been used in LQC, where in order to connect lattice models and symmetry-reduced models one def\/ines averages of f\/luxes over a~lattice, rather than using a total f\/lux (see, e.g., Section~3.6.1 in~\cite{BojOSIGMA}). Also in that context, the number of layers to average over would be $N^{1/3}$. The factor~$N^{1/3}$ is also consistent with a similar construction in~\cite{LQCLQG1}, also rooted in LQC. All this suggests~(\ref{scalefdef}) as the most natural and useful def\/inition of a cosmological scale factor~$a$ from the ``total area''.

The study of inhomogeneities in GFT condensates is still in its infancy, and dif\/ferent ideas for their study have been put forward. One approach presumes that the quantum notion of wavefunction homogeneity does not necessarily lead to a universe that would be observed to be exactly homogeneous. Indeed, in any scenario which is not a truncation to homogeneity but an embedding of homogeneous geometries into a more general framework, one expects inhomogeneities to be present as quantum f\/luctuations, as restricting f\/luctuations over homogeneity to be exactly zero for both metric and connection would seem to violate the Heisenberg uncertainty relations. A prime example of this is inf\/lation, where in the early universe quantum f\/luctuations undergo a transition to classical inhomogeneities, visible as temperature perturbations in the cosmic microwave background (CMB) today \cite{QuantumClassical}, in spite of the initial state being homogeneous. Put dif\/ferently, condensate states have nonvanishing overlap with generic other states in the Fock space, in particular states that one would associate with generic inhomogeneous universes.

\begin{figure}[t]\centering
\includegraphics[scale=0.5]{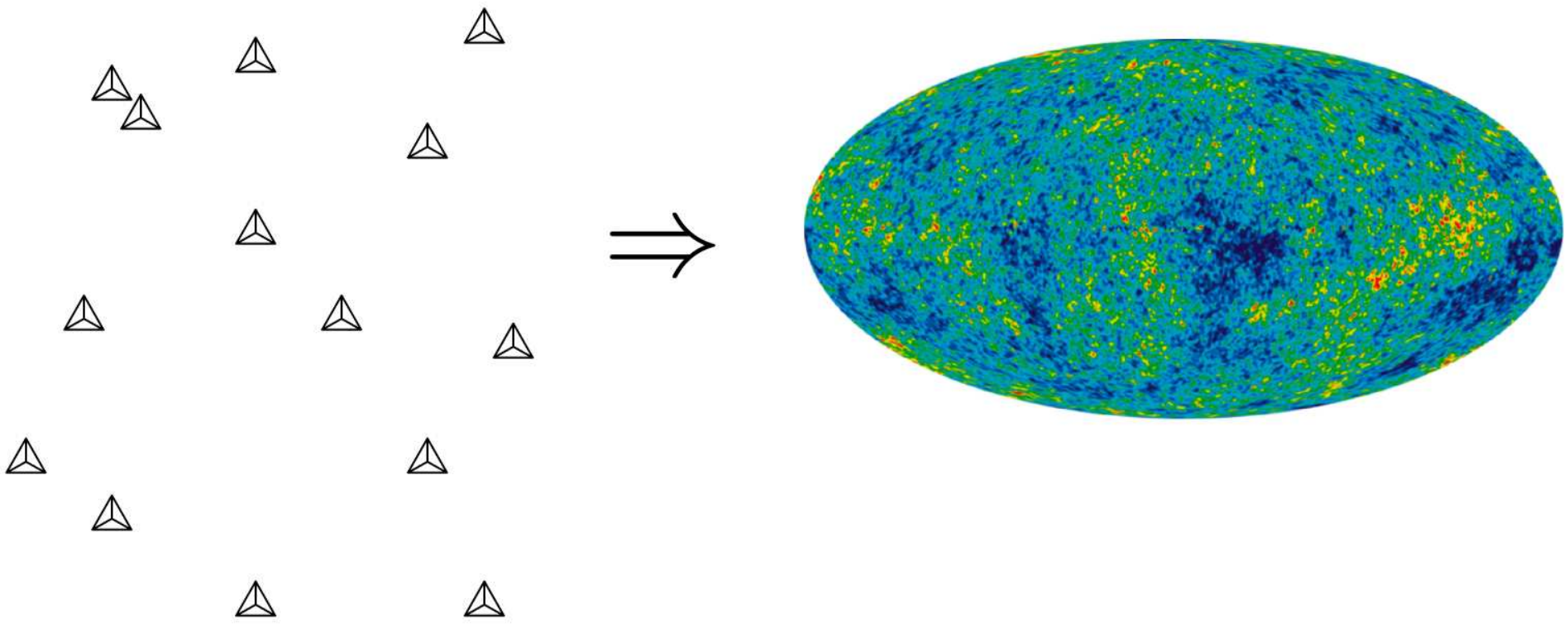}
\caption{One of the main goals of research on GFT condensates.}\label{cmbfig}
\end{figure}

The mean-f\/ield approximation provides a direct way of implementing a transition from quantum to classical inhomogeneities for quantum geometry as represented by a GFT condensate~\cite{SteffenJHEP}. Indeed, as discussed in Section~\ref{wavefsec}, in this mean-f\/ield approximation the ``condensate wavefunction'' represents a classical f\/ield conf\/iguration -- in the case of a real Bose--Einstein condensate it def\/ines the density prof\/ile for a f\/luid in the laboratory. In this sense, in~(\ref{singlecon}) the f\/ield $\sigma(g_I)$ represents a classical f\/ield conf\/iguration on the ``connection minisuperspace'' of group ele\-ments~$g_I$, and its noncommutative Fourier transform $\tilde\sigma(B_I)$ (cf.~(\ref{ncsigma})) a classical f\/ield on the ``metric minisuperspace'' of bivectors~$B_I$. For any subset $C$ of such a minisuperspace, $\int_C |\sigma|^2$ counts the expected number of ``patches'' with metric/connection data in~$C$; in this approximation, all ``patches'' have the same probability distribution of geometric data but they need not all be in exactly the same conf\/iguration when ``measured'' at a given moment. One distinguishes between wavefunction homogeneity as introduced in Section~\ref{simplecondsec} and statistical, classical homogeneity, and is led to a picture of a ``many-patch model'' with a statistical distribution of inhomogeneities, just as is observed in the CMB, see Fig.~\ref{cmbfig}. Some information about this statistical distribution can be extracted from expectation values of more complicated operators such as
\begin{gather*}
\widehat{\alpha}_I = \kappa^2 \int \dd g\, \hat\varphi^\dagger(g_J)  (-\Delta_{g_I})\hat\varphi(g_J),
\end{gather*}
by comparing them with the area $\langle\widehat{A}_I\rangle$. As argued in \cite{SteffenJHEP}, in the case of an isotropic geometry, and assuming a single scalar perturbation $\psi(\vec{x})$, one can approximately identify
\begin{gather}\label{integral}
\int_{\mathbb{R}^3} \dd x\,|\psi(\vec{x})|^2 \approx \frac{V_0}{4}\left(\frac{\langle\widehat{\alpha}_I\rangle N}{\langle \widehat{A}_I\rangle^2}-1\right),
\end{gather}
which reduces to zero in the exactly (quantum) homogeneous case where $\langle\widehat{\alpha}_I\rangle N=\langle \widehat{A}_I\rangle^2$.

Hence, even without any information about the embedding of the patches into space, limited information about perturbations can be extracted. The results of~\cite{SteffenJHEP} outline the potential for a systematic analysis of classical inhomogeneities emerging from quantum f\/luctuations in the mean-f\/ield approximation. For more complicated condensate states, the precise relation is much less clear, and deserves further work.

An essentially orthogonal point of view, though not incompatible with the previous one, is that inhomogeneities should be included as f\/luctuations over the exact condensate, i.e., as phonons~\cite{JHEP}, or more simply as generalisations of the state~(\ref{singlecon})~\cite{SteffenPRD}. While looking at such perturbed condensates is no problem in principle, one faces the problem of interpreting the background-independent geometric data associated to the perturbation, given by group or Lie algebra elements, in terms of continuum f\/ields on a manifold, where the manifold and the homogeneous background itself should be formed by the unperturbed condensate. One expects some similarities with the Born--Oppenheimer approximation in usual quantum cosmology (see, e.g.,~\cite{kiefer1,kiefer2}) in the separation of degrees of freedom into homogeneous and inhomogeneous parts, where the ``background'' is formed by a semiclassical wavefunction for the homogeneous mode. An interesting result of~\cite{SteffenPRD} was that the ratio of the number of quanta in the homogeneous background and in the f\/luctuation over it provides a natural small parameter not present in the continuum. However, what is missing so far is a notion of separating perturbation modes by ``wavenumber'', i.e., a Laplacian or spectral decomposition of modes with respect to the background geometry (presumably unrelated to the Laplacian on the group $G$ used in studies of GFT renormalisation~\cite{GFTrenorm3,GFTrenorm4,GFTrenorm5}). The perturbations considered in~\cite{SteffenPRD} appear to be merely perturbations in the homogeneous mode, not qualitatively dif\/ferent from the unperturbed condensate.

\subsection{Discreteness and continuum limit}\label{contlsec}

The geometries def\/ined by $N$-particle states in GFT or equivalently by a f\/ixed graph in LQG are clearly discrete. In order for them to admit even an approximate continuum interpretation, $N$ must be very large, and the curvature should be very small when compared to the length scale of a tetrahedron. So far, there is no clear def\/inition of a curvature operator in the GFT setting, which would be needed, through an analysis of its expectation values, to make this statement fully rigorous. In the more specif\/ic setting of condensate states, it is possible to still make this idea more precise by only referring to homogeneous degrees of freedom, as has been done in~\cite{JHEP}; for instance, a near-f\/latness condition can be imposed by demanding that the average value of a function $\chi(g_ig_4^{-1}g_jg_4^{-1})$, where $\chi$ is a character on the group, e.g., the trace in the $j=\half$ representation for $G=\SU$, is very close to $\chi(e)$ for all $i,j$. For a more detailed analysis, in a~specif\/ic example, of condensate wavefunctions that correspond to such conf\/igurations of nearly f\/lat tetrahedra, and hence admit an approximate continuum picture, see \cite{SteffenCQG}.

As we have mentioned, $N\rightarrow\infty$ corresponds to a continuum limit, which has to be understood as a phase transition,
as it has been shown in great detail in the case of matrix models for two-dimensional gravity \cite{diFrancesco}.
For a phase transition to happen, the coupling constants of the GFT action will have to be tuned
to critical values (if they exist) at which the expectation value of the number of quanta
contained in the physical vacuum
diverges. Phase transitions are only strictly possible for an inf\/inite number of degrees of freedom, which means one is often mainly interested in the limit $N\rightarrow \infty$. Simultaneously, this ``inf\/inite volume'' limit has to be accompanied by
the divergence of correlation lengths of matter f\/ields used to probe spacetime geometry. It is only in the continuum limit that symmetries such as dif\/feomorphism symmetry can be restored \cite{Biancarev}.

On the other hand, one of the central results of LQG has been the realisation that, at least at the kinematical level, the spectra of area and volume operators are discrete~\cite{volume2}, although whether this implies discreteness for physical states is far from clear \cite{BiancaThomas}. If there is a physical discreteness in the theory, we would be in a situation similar to causal sets \cite{CSTreview} where there is no continuum limit but only a ``continuum approximation''. A~GFT condensate describing a~universe similar to ours would then have an extremely large but still f\/inite number of degrees of freedom~$N$, which for practical purposes may still be approximated as inf\/inite.

\looseness=1 The formalism of GFT condensates for cosmology, in the way it has been set up so far, essentially requires~$N$ to be f\/inite. First of all, this is because the limit $N\rightarrow\infty$ corresponds to a change of representation to a dif\/ferent Hilbert space, and while such a change may be highly desirable eventually to describe a phase transition, we are performing all calculations in the original Fock space around the perturbative GFT vacuum $\varphi=0$. Furthermore, quantities of cosmological interest such as the scale factor~$a$ in~(\ref{scalefdef}) are def\/ined using~$N$, and become ill-def\/ined in the $N\rightarrow\infty$ limit; we are def\/ining approximate continuum quantities from the discrete variables of GFT, before any double-scaling limit of the usual type (discretisation scales go to zero, quantities of physical interest remain f\/inite) is performed. In order to extend these constructions to the limiting case where the actual continuum limit is taken, a~better understanding of this limit in GFT and the change in Fock space representation will be required. This said, it is also true that coherent states, even if they have a f\/inite expectation value of the number operator, have nonvanishing projection over states with arbitrarily large number of quanta. Therefore, these approximate states are taking into account the idea of a~continuum limit, even if in this limited sense. This is perhaps the most signif\/icant dif\/ference with respect to states built with a given number of degrees of freedom attached to a f\/ixed graph.

Cosmologies based on a universe with a f\/inite number of discrete degrees of freedom can have interesting properties of immediate phenomenological relevance. Perhaps the most striking is a~proposal in the causal set approach~\cite{sorkinlambda}, in which the smallness of the cosmological constant $\Lambda$ is conjectured to be a statistical ef\/fect arising from a very large but f\/inite number of constituents of the universe. One expects f\/luctuations in $\Lambda$ to be of the order $\Lambda\sim 1/\sqrt{N}$, and if~$N$ is taken to be the number of Planck volumes in a Hubble volume this turns out to be roughly the observed value (this property is often considered part of the cosmological constant problem, the ``coincidence problem''). While more of a general argument than a strict derivation, this idea shows how properties of the universe at the largest scales might depend on a fundamental discreteness of spacetime. More speculative ideas include attempts to derive deviations from scale invariance in the CMB power spectrum from fundamental discreteness \cite{dreyer}.

The formalism of LQC is based on fundamental discreteness in an essential way. Indeed, its departures from the classical continuum Friedmann equation or Wheeler--DeWitt quantum cosmology arise from using the input of a non-zero minimal area from the kinematics of LQG \cite{LQCrev1,LQCrev2}. In the derivation of the modif\/ications to the continuum dynamics, the picture of a~cosmological universe as def\/ined on a large but f\/ixed graph is used rather explicitly, e.g., in~\cite{improdyn}, without using a full LQG calculation. A crucial question in this context is how the (locally f\/inite) number of degrees of freedom evolves with the expansion of the universe. The viewpoint emerging from \cite{improdyn} was that the volume of (a comoving portion of) the universe and the number of fundamental degrees of freedom should essentially be proportional, so that the volume per elementary cell is a constant (given by its minimal, Planckian value). More generally, LQC suggests a {\em lattice refinement} interpretation in which both quantities could evolve as the graph is ref\/ined in the evolution of the universe \cite{latticerefine}. We will see later on (Section~\ref{dynsec}) that the picture of lattice ref\/inement and the resulting form of holonomy corrections in LQC can be grounded in a natural way in the formalism of GFT condensates.

The picture of a graph to which new degrees of freedom are continuously added as the universe evolves leads to many open questions. One needs to ensure that any notion of homogeneity one wants to apply is preserved under such an evolution, which itself is fundamentally irreversible. For some work on this type of quantum dynamics, using Hilbert spaces of evolving dimension, see~\cite{philipp}. The open issue of how lattice ref\/inement, as apparently required in LQC, is implemented dynamically in LQG provides additional motivation for the study of GFT condensates and their ef\/fective dynamics, which may provide a link between LQG and LQC by dynamically realising lattice ref\/inement. Note that lattice ref\/inement requires a~dynamically changing number of degrees of freedom~$N$, which indicates that working on a f\/ixed graph in LQG may not be the appropriate setting for cosmology.

\subsection{Generalised condensates and generalisation to spherical symmetry}\label{gencond}

As discussed in Section~\ref{simplecondsec}, the simplest condensate states of the form (\ref{singlecon}) or (\ref{dipolecon}), while ha\-ving attractive mathematical properties and a clear motivation by analogy with Bose--Einstein condensates, have shortcomings in their interpretation in terms of quantum geometry. In particular, we have seen that the absence of a connected graph means that there is no information about the topology of space, and that it is dif\/f\/icult to localise excitations, as seems required if one wants to explicitly add perturbations. This suggests def\/ining {\em generalised condensates} on a nontrivial, connected graph. Here we review the essential ideas and technical points of this def\/inition; full details can be found in~\cite{GeneralCondensates}.

Generalised condensates, just as the simple condensates of Section~\ref{simplecondsec}, are based on the idea of wavefunction homogeneity, i.e., the property of a many-particle state to be fully determined by a single-particle wavefunction. Denote this wavefunction again by $\sigma$ and def\/ine operators
\begin{gather*}
\hat\sigma(h_I)=\int \dd g \,\sigma(h_Ig_I)\hat\varphi(g_I),\qquad \hat\sigma^\dagger(h_I)=\int \dd g \,\bar\sigma(h_Ig_I)\hat\varphi^\dagger(g_I).
\end{gather*}
Now f\/ix a graph $\Gamma$ which encodes the desired topology, in the sense that it corresponds to the dual of a triangulation of a manifold of this topology, and def\/ine the ``seed state''
\begin{gather*}
|{\rm seed}\rangle\propto\int (\dd h)^{\# V(\Gamma)} \prod_{e\in E(\Gamma)}\delta\big(h^v_I (h^w_J)^{-1}\big) \prod_{u\in V(\Gamma)}\hat\sigma^\dagger(h^u_I)|0\rangle,
\end{gather*}
where $V(\Gamma)$ denotes the vertices in $\Gamma$ and $E(\Gamma)$ the edges of $\Gamma$; again we are using a slight shorthand notation where each edge~$e$ is characterised as outgoing as the $I$-th link from the vertex $v$ and incoming as the $J$-th link into the vertex~$w$.

The ``seed state'' is of the type often used in LQG, a state on a f\/ixed graph, with a small number of degrees of freedom with respect to the Ashtekar--Lewandowski vacuum. To def\/ine from it a~type of condensate representing some sort of continuum limit, and to incorporate a~notion of coarse graining related to ideas such as \cite{bianca2}, one now def\/ines {\em refinement operators} which preserve the topology of the graph and the notion of wavefunction homogeneity. Both properties can be implemented by basing the ref\/inement operators on Pachner moves~\cite{pachner}, and by def\/ining them using the wavefunction $\sigma$ that was already used for the seed state. For instance, in order to implement a $1\rightarrow 4$ Pachner move one should f\/ind an operator $\widehat{\mathcal{M}}$ satisfying~\cite{GeneralCondensates}
\begin{gather*}
\big[\widehat{\mathcal{M}},\hat\sigma^\dagger(h_1,h_2,h_3,h_4)\big] = \int_{G^6} \dd g\, \left(\hat\sigma^\dagger(h_4,g_1,g_2,g_3)\,\hat\sigma^\dagger(g_3,h_3,g_4,g_5)\right.\nonumber\\
\left. \hphantom{\big[\widehat{\mathcal{M}},\hat\sigma^\dagger(h_1,h_2,h_3,h_4)\big] =}{}\times
\hat\sigma^\dagger(g_5,g_2,h_2,g_6) \hat\sigma^\dagger(g_6,g_4,g_1,h_1)\right),
\end{gather*}
corresponding to the addition of three new vertices and six new edges to the graph; in this move, a tetrahedron is replaced by a more ref\/ined triangulation of a 3-ball, homeomorphic to the tetrahedron that was removed. Crucially, the action of $\widehat{\mathcal{M}}$ does not create disconnected components but only ref\/ines the given graph.

Once such an operator has been found, one can def\/ine generalised condensate states
\begin{gather}\label{gencon}
|\Psi_F\rangle = F\big(\widehat{\mathcal{M}}\big)|{\rm seed}\rangle
\end{gather}
for some function $F$. A natural choice motivated by technical simplicity and the analogy to the simple condensates of Section~\ref{simplecondsec} would be an exponential. Even for this choice, expectation values for operators of geometric interest are quite dif\/f\/icult to compute. An interesting general result shown in~\cite{GeneralCondensates} is that, for a certain class of ref\/inement moves and initial seed states, expectation values of one-body operators of the general form
\begin{gather*}
\widehat{\mathcal{A}}=\int \dd g\,\dd h\, A(g_I,h_I) \hat\varphi^\dagger(g_I)\hat\varphi(h_I)
\end{gather*}
are extensive, i.e., reduce to a product of the average particle number $N=\langle\widehat{N}\rangle$ and a single-particle expectation value given in terms of the wavefunction $\sigma$. This result is important for cosmology, where the scaling of observables with~$N$ is a main property of interest.

The construction of generalised condensates leads in a rather straightforward way to a notion of spherical symmetry. The seed state can be chosen to correspond to a topology $S^2\times[0,1]$, interpreted as a spherical shell, where one needs to def\/ine dif\/ferent types of vertices that keep track of the inner boundary, outer boundary, and interior of the shell~\cite{GeneralCondensates}. Gluing many such shells together using the inner boundary of one and the outer boundary of the next one, one obtains a more general conf\/iguration of spherical symmetry with nontrivial dependence along the radial direction, given by dif\/ferent wavefunctions $\sigma_r$ where $r$ labels the dif\/ferent shells.

For the construction of generalised condensates, it has proven advantageous to make use of tools developed in the context of coloured tensor models
(see, e.g., \cite[Section~4]{GurauRyan} for a concise presentation of the relevant procedures). The simple GFT models based on a single f\/ield are replaced with coloured models, possessing additional labels that facilitate the reconstruction of the topological structure of the boundary state, given the pattern of convolutions of the wavefunction. While the additional labels do not have an obvious counterpart in LQG, they are useful to deal with the problem of isolating the various contributions to quantities computed for families of dif\/ferent graphs. As the particular colouring introduced requires bipartite graphs, Pachner
moves have to be replaced with \emph{melonic refinement moves}. These ref\/inement moves allow a better handling of the construction of a~ref\/ined foliation, keeping f\/ixed the topology. These small extensions are only required to make the construction easier to be dealt with in concrete cases, and do not change the substance of the general principles that we have discussed.

\looseness=-1 A f\/irst very interesting application of this formalism has been given in \cite{HorizonEntropy} where it was shown how to def\/ine a horizon entanglement entropy for such generalised condensates. There, thanks to the techniques developed in the second quantised formulation of GFT and of tensor models, it has been possible to get a~f\/irst hint of the ef\/fect of superposing dif\/ferent graphs, a~crucial aspect of the condensate picture, on a key physical quantity for semiclassical gravity, namely the entanglement entropy of subregions. It was shown that the entanglement entropy associated to an isolated horizon~\cite{IsolatedHorizons}, def\/ined in terms of GFT condensates, receives a (combinatorial) Boltzmann entropy contribution from the structure of the state in terms of its graph components, as the graph proliferation associated to the ref\/inement moves involved has the correct scaling with the number of tetrahedra in the state. This feature of the generalised coherent states, in turn, gives rise to an area law with subleading logarithmic corrections, as expected from semiclassical gravity. This result already underscores the usefulness of GFT condensates and tensor model techniques for calculations that address the physics of many GFT degrees of freedom.

\section{GFT condensates: dynamics}\label{dynsec}

In Section~\ref{kinesec}, we described the construction of GFT condensate states at the kinematical level, and given their possible interpretation in terms of geometry and cosmology. The next step is to study the ef\/fective dynamics of such states, as extracted from the quantum dynamics of a~given GFT model or class of GFT models. The restriction to a class of states such as~(\ref{singlecon}) or~(\ref{gencon}) can be suf\/f\/icient for obtaining manageable equations, but often further approximations are needed. Showing that there is a relation between the resulting ef\/fective dynamics and the dynamics of cosmological spacetimes in general relativity, such as the Friedmann equation for FRW backgrounds or dynamical equations for linearised perturbations, would not only give further support for the geometric interpretation of the abstract group-theoretic data appearing in GFT that we have put forward, but more importantly also give evidence that a certain GFT model, or class of models, can become serious contenders for a theory of quantum gravity in four dimensions. In this vein, one can hope to use GFT condensates to discriminate between dif\/ferent models, by restricting only to those that lead to an acceptable cosmological phenomenology. Ideally, a consistent picture of geometrogenesis would not only lead to a resolution of the Big Bang singularity in a consistent quantum theory, but also obviate the need for introducing inf\/lationary models with their known theoretical problems~\cite{inflation}.

As we will show in this section, the basic strategy for deriving ef\/fective dynamics for condensates from the fundamental GFT dynamics is rather straightforward, and in many ways analogous to how one proceeds in the case of Bose--Einstein condensates in condensed matter physics. We will give some detail on the resulting equations, while trying to be as general as possible, given the multitude of models in the GFT literature.

It has proven more dif\/f\/icult to identify interesting solutions to the resulting equations and to map them to an ef\/fective classical or semiclassical cosmological description. This is mainly because the resulting equations are almost always nonlinear and nonlocal with a peculiar type of nonlocality, and not many analytical solution techniques for such equations are known. Numerical techniques, while certainly a possible way forward, have not been employed much, and most of the work so far has focused on general and foundational aspects of the formalism.

\subsection{Gross--Pitaevskii equation as ef\/fective dynamics of a Bose condensate}

In the interest of making this review self-contained for readers who are not familiar with quantum f\/ield theory techniques in condensed matter physics, let us brief\/ly review how the Gross--Pitaevskii equation, a certain type of ``nonlinear Schr\"odinger equation'', emerges from the ef\/fective dynamics of a~quantum f\/ield theory for the atoms of a Bose--Einstein condensate.

In the simplest example, this quantum f\/ield theory can be def\/ined by an action
\begin{gather*}
S[\Psi,\bar\Psi]=\int_{\mathbb{R}^3\times\mathbb{R}}\! \dd x\,\dd t\left\{\frac{\im\hbar}{2}\left[\bar\Psi\frac{\partial\Psi}{\partial t}-\Psi\frac{\partial\bar\Psi}{\partial t}\right]-\left(\frac{\hbar^2}{2m}|\vec\nabla\Psi|^2-\mu|\Psi|^2+V(\vec{x})|\Psi|^2+\frac{\lambda}{2}|\Psi|^4\right)\!\right\},
\end{gather*}
where the scalar f\/ield $\Psi$ and its complex conjugate $\bar\Psi$, functions on non-relativistic f\/lat spacetime $\mathbb{R}^3\times\mathbb{R}$, are conventionally treated as independent in the variation (this is basically equivalent to splitting the complex f\/ield $\Psi$ into two real components). Here $\mu$ represents a chemical potential and $\lambda$ is a coupling representing the (weak) interactions between atoms. We assume that the atoms have zero spin and that there is only a single species, but generalisations to spin degrees of freedom and to several species are straightforward. More importantly, it is assumed that the condensate is examined in a dilute limit, in which the interatomic distance is larger than the range of the interactions. This action should be seen as the f\/irst approximation of a more complicated theory which will include more terms describing all the possible interaction channels between the atoms. It is an ef\/fective f\/ield theory for the order parameter controlling a phase transition, whose regime of validity has to be checked a posteriori. However, it is also a generic prototype for a variety of dif\/ferent situations in which a weak coupling approximation, related to the low density of the f\/luid, can be applied.

Because the action is real, variations with respect to $\Psi$ and $\bar\Psi$ lead to only one independent complex equation, known as the {\em Gross--Pitaevskii equation},{\samepage
\begin{gather}\label{becfieldeq}
\im\hbar\frac{\partial\Psi}{\partial t} = -\frac{\hbar^2}{2m}\Delta\Psi -\mu\Psi+V(\vec{x})\Psi+\lambda|\Psi|^2\Psi.
\end{gather}
This is the classical f\/ield equation for a complex f\/ield $\Psi$, nonlinear in $\Psi$ due to the last term.}

Proceeding with canonical quantisation, the f\/irst term in the action tells us that $\Psi$ and $\bar\Psi$ are canonically conjugate, and they should satisfy the Heisenberg algebra as operators, analogous to the GFT commutation relations~(\ref{commutation}). A~Fock space is constructed in which $\hat\Psi(\vec{x})$ is a sum of annihilation operators and $\hat\Psi^\dagger(\vec{x})$ is a~sum of creation operators with respect to a Fock vacuum~$|0\rangle$. The excitations created by the ladder operators represent atoms.

The quantum dynamics given by the operator version of (\ref{becfieldeq}) is nonlinear and explicit solutions are hard to obtain in general. To proceed, one uses, in the simplest case, a {\em mean-field approximation} in which the operators are replaced by classical f\/ields, i.e.,
\begin{gather*}
\hat\Psi(\vec{x}) = \psi(\vec{x}) {\bf 1}+\hat{\chi}(\vec{x}),\qquad \hat\Psi^\dagger(\vec{x}) = \bar\psi(\vec{x}) {\bf 1}+\hat{\chi}^\dagger(\vec{x}),
\end{gather*}
where one assumes that the expectation values of $\hat\chi$ and $\hat\chi^\dagger$ which represent f\/luctuations over the mean f\/ield $\psi$ and $\bar\psi$ are small and can be neglected in the simplest approximation. In such an approximation, the operator version of~(\ref{becfieldeq}) evidently reduces again to the classical f\/ield equation~(\ref{becfieldeq}) with~$\Psi$ and~$\bar\Psi$ replaced by~$\psi$ and~$\bar\psi$.

To make this more precise, we would like to justify the mean-f\/ield approximation in the Fock space formalism by constructing a state satisfying $\hat\Psi(\vec{x})|\psi\rangle = \psi(\vec{x})|\psi\rangle$. This property is satisf\/ied by the {\em coherent state}
\begin{gather*}
|\psi\rangle\propto\exp\left(\int_{\mathbb{R}^3} \dd x\,\psi(\vec{x}) \hat\Psi^\dagger(\vec{x})\right)|0\rangle.
\end{gather*}
The similarity to our construction (\ref{singlecon}) for GFT should be obvious. It is also clear that there is no such state that also satisf\/ies $\hat\Psi^\dagger(\vec{x})|\psi\rangle = \bar\psi(\vec{x})|\psi\rangle$, which would be in contradiction with the fact that $\hat\Psi$ and $\hat\Psi^\dagger$ do not commute. Hence the mean-f\/ield approximation can only correspond to an expectation value of the (normal ordered) operator equations of motion;
\begin{gather}\label{BECexpvalue}
\left\langle\psi\left|\frac{\delta S[\hat\Psi,\hat\Psi^\dagger]}{\delta\hat\Psi^\dagger}\right|\psi\right\rangle=\left\langle\psi\left|\im\hbar\frac{\partial\hat\Psi}{\partial t}+\frac{\hbar^2}{2m}\Delta\hat\Psi +\mu\hat\Psi-V(\vec{x})\hat\Psi-\lambda\hat\Psi^\dagger\hat\Psi^\dagger\hat\Psi\right|\psi\right\rangle=0
\end{gather}
is equivalent to the Gross--Pitaevskii equation for the f\/ield $\psi$.

We repeat that, for the general nonlinear Gross--Pitaevskii equation, there is no consistent interpretation of $\psi(\vec{x})$ as an actual single-particle wavefunction; there is no probability interpretation attached to it, as one has already taken an expectation value, and there is no Hilbert space based on such ``wavefunctions'' which are essentially parameters for a certain class of states in the Hilbert space of the theory, the non-relativistic Fock space. Rather, this wavefunction includes hydrodynamic information, as it can be shown using the Madelung representation (cf.~(\ref{BECPsi})) $\psi(\vec{x}) = \sqrt{\rho(\vec{x})} \exp(-\iu \theta(\vec{x}))$. The modulus of the wavefunction controls the number density, while the phase controls the velocity through the velocity potential $\theta$, such that $\vec{v} = \frac{\hbar}{m}\vec{\nabla}\theta$. See Section~\ref{wavefsec} in the kinematical part for an extended discussion of this point, and its meaning for the ``nonlinear quantum cosmology'' derived from the ef\/fective dynamics of GFT condensates.

When interactions can be ignored ($\lambda=0$), a single-particle wavefunction solving the (then standard) Schr\"odinger equation appears as the function $\psi(\vec{x})$ in the mean-f\/ield approximation; the latter is then simply the statement that one has a large number of identical non-interacting bosonic particles in a certain potential.

The main lesson from the well-understood case of condensed matter physics for our treatment of the ef\/fective GFT dynamics is the importance of approximations not only in the choice of state but also in the truncation of the full dynamics to expectation values such as (\ref{BECexpvalue}). Indeed, in much of the following we will explore the hypothesis that the analogue of (\ref{BECexpvalue}), in the case of GFT, provides a good approximation to the full quantum GFT dynamics.

\subsection{Ef\/fective equations from GFT Schwinger--Dyson equations}

Given that GFTs, as introduced in Section~\ref{gftsec}, are quantum f\/ield theories with a structure rather similar to the non-relativistic f\/ield theories used in condensed matter physics, the most straightforward way of obtaining an ef\/fective dynamics for GFT condensates is from expectation values similar to~(\ref{BECexpvalue}). The simplest choice is indeed
\begin{gather}\label{GFTexpvalue}
\left\langle\sigma\left|\frac{\delta S[\hat\varphi,\hat\varphi^\dagger]}{\delta\hat\varphi^\dagger(g_I)}\right|\sigma\right\rangle=0,
\end{gather}
where $S$ is now the action of a given GFT model and $\hat\varphi$ and $\hat\varphi^\dagger$ are the f\/ield operators introduced in Section~\ref{gftsec}. For the simplest choice of condensate state~(\ref{singlecon}), which implements the mean-f\/ield approximation, this then directly gives the GFT analogue of the Gross--Pitaevskii equation: the mean f\/ield $\sigma(g_I)$ should satisfy the classical GFT equations of motion,
\begin{gather}\label{classicalEOM}
\int \dd h \, K(g_I,h_I)\sigma(h_I)+\frac{\delta V[\sigma,\bar\sigma]}{\delta\bar\sigma(g_I)}=0.
\end{gather}
This equation is now not only nonlinear {\em but also nonlocal} in $\sigma$, even for a local kinetic opera\-tor~$K$, due to the nonlocal combinatorial pairing of group arguments in the interaction part~$V$. This suggests that exact solutions will not be easily obtained.

Already for the ``dipole condensate'' (\ref{dipolecon}) and certainly for the generalised condensates~(\ref{gencon}), the expectation value (\ref{GFTexpvalue}) is much harder to obtain. In general one will not f\/ind a closed expression in terms of the ``condensate wavefunction'' (also denoted $\sigma$ in the general case) without further approximations, such as an expansion in powers of~$\sigma$ that is truncated at some order (see, e.g.,~\cite{LorenzoFidelity}). Certain choices for~$K$ and~$V$ can lead to models that can be solved more directly.

One may also be interested in going beyond the truncation to only a single expectation value~(\ref{GFTexpvalue}). Indeed, (\ref{GFTexpvalue}) can be seen as the simplest of an inf\/inite tower of {\em Schwinger--Dyson equations} for the GFT model in question. Formally, Schwinger--Dyson equations can be derived from assuming a ``fundamental theorem of functional calculus'' by which, schematically,
\begin{gather}\label{SDeq}
0 = \int \mathcal{D}\varphi\,\mathcal{D}\bar\varphi \, \frac{\delta}{\delta\bar\varphi(g_I)}\left(\mathcal{O}[\varphi,\bar\varphi]e^{-S[\varphi,\bar\varphi]}\right) = \left\langle \frac{\delta\mathcal{O}[\varphi,\bar\varphi]}{\delta\bar\varphi(g_I)}-\mathcal{O}[\varphi,\bar\varphi]\frac{\delta S[\varphi,\bar\varphi]}{\delta\bar\varphi(g_I)}\right\rangle
\end{gather}
holds in the vacuum state for any suf\/f\/iciently well-behaved functional $\mathcal{O}$ of the f\/ield and its complex conjugate (such as a polynomial). Since we are looking for condensates that can play the role of a new, non-perturbative vacuum of the theory, one can impose~(\ref{SDeq}), for some choices of~$\mathcal{O}$, as consistency conditions relating dif\/ferent $n$-point functions, which have to be satisf\/ied in a physical condensate state. Using Schwinger--Dyson equations to extract an ef\/fective dynamics of GFT condensates was f\/irst proposed in~\cite{PRL,JHEP}.

In practice, beyond the truncation imposed by considering a few very simple choices for $\mathcal{O}$, so far another important approximation has been made: $n$-point correlation functions are computed as expectation values in the inner product of the Fock space, i.e., of the Hilbert space of a~``free'' GFT. One could try to go beyond this approximation using methods such as used in~\cite{dine}, where general closed expressions for two-point functions were derived using Schwinger--Dyson equations and then solved in a perturbative expansion in the GFT coupling constant(s).

\subsection{Ef\/fective equations from projection methods}
So far we have proceeded via f\/ield-theoretic considerations. We want to brief\/ly mention further possibilities to def\/ine an ef\/fective dynamics from the microscopic theory, introducing some form of coarse graining. In many condensed matter physics problems, in the majority of the cases where the exact ground state is not known, the objective is to f\/ind the best approximation for the properties of the ground state of a system, while keeping track of a reasonable amount of observables. In GFT we do not have a proper notion of energy to minimise, but it is still possible to adapt many of the techniques used in condensed matter physics. In doing so, one can really appreciate the power of the second quantised formulation that we are using.

The ultimate objective of the GFT programme as we have presented it is to provide predictions for physical observables, even in a simplif\/ied context as the one of cosmological dynamics. This means that the f\/inal results will be specif\/ic relations between the observables of relevant physical quantities (in our
case, scale factors, Hubble parameter, etc.), as is also a necessity in a background-independent language as the one of GFT. The spirit of condensate states that we are advocating is clear: they are a way to obtain such a coarse-grained picture by identifying a simple set of variables (contained implicitly in the condensate wavefunction) and restricting their values with some equations of motion appropriately derived from the microscopic theory.

It is natural to ask whether one can bypass the condensate idea and work out the relations between observables directly. A possibility, proposed in~\cite{LorenzoFidelity}, is to adapt the projection methods used in statistical mechanics to derive the equations of motion of selected observables from
the (quantum) Liouville equation for the whole system~\cite{Balian, BreuerPetruccione}.

Instead of working with a pure state, we will have to work with a (generically mixed) density matrix adapted to the relevant observables $\mathfrak{A} = \{ \hat{A}_i, \, i=1,\ldots, N \}$, which is designed using a~maximum entropy principle. The idea is that, at f\/ixed (but unspecif\/ied) values of the expectation values~$\mathcal{A}_i$ of the observables in~$\mathfrak{A}$, the density matrix that maximises the von Neumann entropy will have the general shape
\begin{gather*}
\hat{\rho}(\gamma) = \exp\big( \Psi(\gamma) - \gamma^i \hat{A}_i \big),
\end{gather*}
where $\gamma^i$ are the chemical potentials, dual to the observables, and $\Psi$ is the logarithm of the partition function, up to a sign. The chemical potentials are determined implicitly by
\begin{gather}\label{expvalues}
\mathcal{A}_i = \frac{\partial \Psi}{\partial \gamma^i}
\end{gather}
as usual. The idea is to determine the values of the chemical potentials such that the density matrix that they identify can be used in place of the exact microscopic state, at least as far as the operators in~$\mathfrak{A}$ are concerned.

The problem, then, is to f\/ind the values of $\gamma$ that best ref\/lect the microscopic equation of motion. In the GFT setting, we do not have notions
such as time evolution and a ground state of minimal energy from which we can construct approximation methods (such as, for instance, the variational method in time-independent quantum-mechanical problems). A natural option is to require that the overlap between the reduced density matrix and a microscopic solution of the equations of motion is maximal. This simple idea is based on the notion of \emph{fidelity}, which has been used within the context of quantum phase transitions~\cite{fidelity} to explore their dynamical aspects. This reasoning leads immediately to a \emph{variational principle} for the observables, which follows from maximising the overlap between the approximate density matrix and the density matrix of the exact state $\ket{\Phi}\bra{\Phi}$
\begin{gather}
\exp(\Sigma(\gamma)) = \mathrm{Tr} \big( \hat{\rho}(\gamma) \ket{\Phi}\bra{\Phi} \big) .
\end{gather}
The ef\/fective equations are obtained varying this action with respect to the chemical potentials, and f\/inding its maxima. The solutions will then
lead, via~\eqref{expvalues}, to the desired expectation values that best match the physical state. Additionally, by looking at f\/luctuations
around the mean values, one can provide a f\/irst assessment of the validity of the given approximation, which is one of the crucial aspects when giving predictions for physical observables.

It goes without saying that these expressions are rather formal, and they require appropriate regularisation in order to be used concretely, especially given the possibility that the physical state might not be a normalisable state in the Fock space. Their usefulness is in the more explicit presentation of a path to be followed in developing a thermodynamic approach to GFT, where only macroscopic quantities directly relevant for physics are used, as well as a direct inclusion of a maximum entropy (or maximum ignorance) principle in the derivation of the ef\/fective equations. As we mentioned for the case of generalised condensates, we should expect that the equations for the observables will be very complicated, requiring further simplif\/ications before being reduced to tractable problems. This is of course not a problem faced by GFT alone.

Interestingly, this discussion can be used to motivate, from a dif\/ferent perspective, the Gross--Pitaevskii equation for GFT condensates. Indeed, for a~simple form of the quantum equations of motion (trivial kinetic term and interaction containing only creation operators), one can show that the exact quantum state solving the equations of motion has a very simple, albeit formal, representation of the form of the generalised condensate states,
\begin{gather*}
\ket{\Phi} \sim \exp\big({-} V[\hat{\varphi}^\dagger] \big) \ket{0}.
\end{gather*}
Such an expression is still formal and does not display clearly the geometric properties of the state. Therefore, even if the exact solution is available, we still need some approximation to extract the information it contains, i.e., we need a~state whose properties we can easily understand and that provides the best approximation for the exact state. If we consider then the problem of maximising the f\/idelity of a coherent state on the physical state, along the lines that we have just discussed, we obtain exactly the desired Gross--Pitaevskii equation~\cite{LorenzoFidelity}.

Besides putting the discussion of the GFT condensates under a dif\/ferent light, this point of view shows how GFT can contribute to the concrete realisation of some of the ideas often considered in emergent gravity scenarios~\cite{emergentreview} in a fully background-independent context.

\subsection{Dynamics of cosmological observables}\label{dynobssec}

Once ef\/fective dynamical equations for GFT condensates have been obtained, from Schwinger--Dyson equations or by projection methods, one needs to connect them to cosmology. This connection is made by identifying cosmological observables as expectation values of GFT Fock space operators, as we have discussed in Section~\ref{interpretsec}. In the simplest case, one is interested in observables that can be constructed out of local variables at the level of each tetrahedron, such as the ``total areas''
\begin{gather*}
A_I=\langle\widehat{A}_I\rangle := \kappa\left\langle \int \dd g\, \hat\varphi^\dagger(g_J) \sqrt{-\Delta_{g_I}}\hat\varphi(g_J)\right\rangle
\end{gather*}
or the ef\/fective metric components (\ref{metriccomp}). It is also possible to construct a notion of curvature that is local for each tetrahedron, as mentioned in Section~\ref{contlsec}. By construction, any quantities of this type, arising as expectation values of one-body operators of the form
\begin{gather}\label{1bodyop}
\widehat{X} = \int \dd g\, \dd g' \, X(g_I,g'_I) \hphid(g_I) \hphi(g'_I),
\end{gather}
are extensive, i.e., scale with the number of quanta~$N$. In order to obtain observables that have the correct large-scale geometric interpretation, expectation values of one-body operators~(\ref{1bodyop}) must be rescaled with appropriate powers of~$N$; for instance, in order to have a notion of curvature, expectation values of one-body operators~(\ref{1bodyop}) must be rescaled with $1/N$~\cite{NJP}.

Ef\/fective dynamical equations can then be translated into equations for the dynamics of observables. To clarify this procedure, let us detail a simple example, building on an observation f\/irst made in \cite{PRL}: choosing $\mathcal{O}=\varphi(g'_I)$ in~(\ref{SDeq}) leads to the consistency condition
\begin{gather}\label{SDeq2}
\int \dd h \, K(g_I,h_I)\left\langle\hat\varphi(g'_I)\hat\varphi(h_I)\right\rangle+\left\langle:\frac{\delta V[\hat\varphi,\hat\varphi^\dagger]}{\delta\hat\varphi^\dagger(g_I)}\hat\varphi(g'_I):\right\rangle=0,
\end{gather}
where we are expressing expectation values in the operator formalism using normal ordering, i.e., the second expectation value is to be taken after normal ordering. An equivalent equation holds for $\mathcal{O}=\bar\varphi(g'_I)$ as there is no contribution from $\delta\bar\varphi/\delta\bar\varphi$ which is zero after normal ordering (it is a delta distribution and hence equal to a fundamental commutator).

One can now assume the dipole condensate state (\ref{dipolecon}) in which all odd $n$-point functions vanish, as is straightforward to verify. Then, in a~GFT model of the form (\ref{simpleaction}) corresponding to one of the usual spin foam models in four dimensions, where the potential is built out of~$\varphi^5$ terms, the second term in (\ref{SDeq2}) only contains odd $n$-point functions and is therefore zero, meaning we are left with two equations for two-point functions,
\begin{gather}\label{effxieq}
\int \dd h \, K(g_I,h_I)\big\langle\xi\big|\hat\varphi^\dagger(g'_I)\hat\varphi(h_I)\big|\xi\big\rangle=\int \dd h \, K(g_I,h_I)\left\langle\xi\left|\hat\varphi(g'_I)\hat\varphi(h_I)\right|\xi\right\rangle=0.
\end{gather}
Thus, in this special case, only the two-point functions of the condensate state and only the ``kinetic'' (quadratic) part of the GFT dynamics contribute to the ef\/fective dynamics. This is clearly a drastic approximation, due to the specif\/ic choice of state (\ref{dipolecon}), and only the simplest of an in principle inf\/inite tower of consistency conditions arising from~(\ref{SDeq}). It would suggest that the precise form of the potential $V$ is not relevant in the ef\/fective description, as long as it is built out of terms with odd powers in the GFT f\/ield. Nevertheless, we can use~(\ref{effxieq}) as a starting point to understand the more general procedure. One can motivate~(\ref{effxieq}) without using the specif\/ic dipole condensate state~(\ref{dipolecon}), by working in a weak-coupling limit of the GFT in which contributions of the interaction terms may be neglected on more general grounds.

Further specifying to a kinetic term that is local, $K(g_I,h_I)=\delta(g_I^{-1}h_I)\mathcal{D}_g$ where $\mathcal{D}$ can be a~dif\/ferential operator, setting $g'_I=g_I$ and integrating, the f\/irst equation of~(\ref{effxieq}) yields
\begin{gather}\label{effexpval}
\int \dd g \big\langle\xi\big|\hat\varphi^\dagger(g_I)\mathcal{D}_g\hat\varphi(g_I)\big|\xi\big\rangle=0,
\end{gather}
which is now of the form of an expectation value for a one-body operator, specif\/ied by $\mathcal{D}_g$, which can be translated into cosmological observables. This very direct route from Schwinger--Dyson equations to cosmological observables, which avoids the need for discussing an ef\/fective Wheeler--DeWitt equation, was f\/irst used in~\cite{NJP}.

In the case $\mathcal{D}_g\propto{\bf 1}$, as is often the starting point for GFT models corresponding to spin foams, (\ref{effexpval}) simply gives $N=0$, i.e., the only solution is the Fock vacuum; for nontrivial solutions to arise, the operator~$\mathcal{D}_g$ should have a nontrivial kernel. Work in GFT renormalisation \cite{GFTrenorm3,GFTrenorm4,GFTrenorm5} suggests that the kinetic term must contain a Laplace--Beltrami operator on the group in order to allow a separation of scales (given by the eigenvalues of this Laplacian), and be of the form
\begin{gather}\label{funnykinetic}
\mathcal{D}_g = -\sum_I \Delta_{g_I} - \mu^2,
\end{gather}
where $\Delta_{g_I}$ is a Laplace--Beltrami operator acting on the $I$-th argument, and $\mu^2$ is a (dimensionless) coupling constant specifying the relative weight between the two terms. With~(\ref{funnykinetic}), the interpretation of~(\ref{effexpval}) given in~\cite{SteffenJHEP} was, for an exactly homogeneous and isotropic universe,
\begin{gather}\label{improder}
a^3 V_0 = \left(\frac{\kappa\mu}{2}\right)^{3/2} N,
\end{gather}
where $a$ is the cosmological scale factor, $V_0$ is a f\/iducial coordinate volume (so that $a^3V_0$ corresponds to the physical volume), and $\kappa$ is the ``Planck area'' of the model. This is a nontrivial result, as it suggests that the physical volume should be proportional to the number of degrees of freedom, as seems required to reproduce the {\em improved dynamics} scheme of LQC~\cite{improdyn} (see also~\cite{latticerefine,LQCfromGFT,NJP} for related discussions of this connection, in particular~\cite{LQCfromGFT} for an alternative derivation of LQC improved dynamics from GFT). There are corrections to~(\ref{improder}) if inhomogeneities are present, see~\cite{SteffenJHEP} for details.

A recent idea that makes a more direct connection to quantum cosmology has been the introduction of {\em relational} cosmological observables in~\cite{VolumeDynamics}. Working in a model in which a~massless scalar f\/ield $\phi$ is introduced explicitly as an argument of the GFT f\/ield,
\begin{gather*}
\hat\varphi(g_I) \rightarrow\hat\varphi(g_I,\phi),
\end{gather*}
one can def\/ine quantities like the number operator at a f\/ixed value of $\phi$,
\begin{gather*}
\hat{N}(\phi_0)=\int \dd g\, \hat\varphi^\dagger(g_I,\phi_0)\hat\varphi(g_I,\phi_0),
\end{gather*}
where $\phi$ plays, as usual in quantum cosmology, the role of a relational clock~\cite{oldQCreview}. In a similar way, one can introduce an operator $\hat{V}(\phi_0)$ for the total volume of the universe at a f\/ixed value of the scalar f\/ield, which is one of the main relational observables of interest in LQC~\cite{LQCrev1,LQCrev2}. One can then obtain an ef\/fective Friedmann equation as a relation between $V(\phi)\equiv\langle\hat{V}(\phi)\rangle$ and its derivatives with respect to $\phi$, and show how it reduces to the Friedmann equations for general relativity and LQC in certain cases; see Section~\ref{LQCrelation} for a summary of these most recent, particularly promising results.

\subsection{Ef\/fective Friedmann equations and relation to LQC}\label{LQCrelation}

\looseness=-1 One of the main goals of the study of GFT condensates has been the derivation of an ef\/fective quantum cosmology dynamics that can be compared with the results of minisuperspace models such as LQC. As discussed in Section~\ref{wavefsec}, such a comparison cannot happen directly at the level of Wheeler--DeWitt-like equations, given that ``condensate wavefunctions'' do not play the role of ``wavefunctions of the universe''. While early work such as~\cite{PRL,JHEP} tried to interpret equations such as~(\ref{effxieq}) directly as wave equations for a quantum cosmology wavefunction and accordingly applied approximation methods such as a WKB approximation to their study, this viewpoint has now been replaced by the study of expectation values that can be associated with cosmological observables. The results obtained by these two approaches are related, given that cosmological observables are often extensive so that there is an appropriate scaling with the number of quanta~$N$ that relates the ``single-particle wavefunction'' and ``condensate wavefunction'' viewpoints; there is no discussion of this scaling in~\cite{PRL,JHEP} which was better understood in~\cite{NJP}.

The appearance of $N$, which has no analogue in standard quantum cosmology or in a~conti\-nuum setting, can allow a derivation of properties of ef\/fective quantum cosmology models from the more fundamental setting of GFT. For example, as shown in \cite{NJP}, the form of {\em holonomy corrections} of LQC can be obtained from the appropriate scaling with $N$ in the cosmological interpretation of GFT observables: parametrising $\SU$ group elements by the coordinates
\begin{gather*}
g=:\sqrt{1-|\vec\pi[g]|^2} {\bf 1}-\im\vec\sigma\cdot\vec\pi[g],\qquad |\vec\pi[g]|\le 1,
\end{gather*}
if such a group element is identif\/ied, for example, with a parallel transport along an edge of coordinate length $l$ in the $x$-direction, we have
\begin{gather*}
g=:\mathcal{P}\exp\int_e\omega \approx \exp(l \omega_x) \quad \leadsto  \quad \vec\pi[g]=-\vec\omega_x \sin(l|\vec\omega_x|)/|\vec\omega_x|,
\end{gather*}
where we use $\mathfrak{su}(2)\simeq\mathbb{R}^3$ to view the connection $\omega$ (assumed to be constant over this edge) as an element of $\mathbb{R}^3$. Note that the introduction of a coordinate system led to the replacement of the coordinate-independent expression $\int_e\omega$ by $l\,\omega_x$ where $\omega=\omega_x dx + \cdots$.  If we then def\/ine a~macroscopic ``total group element'' operator by
\begin{gather*}
\widehat{\Pi_i^2}:=\int \dd g\, \big(\vec{\pi}[g_ig_4^{-1}]\big)^2 \hat\varphi^\dagger(g_I) \hat\varphi(g_I),
\end{gather*}
the results of \cite{SteffenJHEP,NJP} show that $l\sim N^{-1/3}$, and thus
\begin{gather}\label{piholo}
\big\langle\widehat{\Pi_i^2}\big\rangle =: N  \sin^2\big((V_0/N)^{1/3} |\vec\omega_i|\big)
\end{gather}
def\/ines an ef\/fective macroscopic connection variable $\omega$. This result shows how the fundamental group variables are related to the sine of the connection multiplied by a function of $N$ that determines the type of holonomy corrections in LQC \cite{LQCrev1,LQCrev2}. In particular, the (tentative) result~(\ref{improder}) would suggest that this function of $N$ is proportional to the inverse scale factor, which precisely def\/ines the improved dynamics scheme in LQC~\cite{improdyn}. The demonstration that the form of LQC holonomy corrections can be obtained from considerations in GFT has been a major achievement of the work on GFT condensates. Note that it requires no assumptions about the specif\/ic form of the dynamics, as it is done essentially at the kinematical level.

While such kinematical results are very promising, one would also like to match the ef\/fective dynamics arising from GFT condensates to a type of Friedmann equation coming from classical general relativity (or a modif\/ied classical theory of gravity) or from a quantum gravity-inspired setting such as LQC. As we have explained, the most promising approach towards this seems to be to compute the dynamics of cosmological observables as expectation values of operators on the GFT Fock space, and to compare the resulting equations with classical or semiclassical dynamical equations for the corresponding quantities. In the setting of LQC, this means focusing on {\em effective equations} that approximate the dynamics for appropriate semiclassical states (see, e.g.,~\cite{LQCrev2}), rather than the full quantum formalism in terms of a Hilbert space and corresponding Wheeler--DeWitt equation. Again, this is consistent with the condensate viewpoint, which suggests that one should work with a special class of semiclassical states, rather than the highly non-classical states given by generic elements of a quantum cosmology Hilbert space. The results of \cite{VolumeDynamics,VolumeDynamicsShort} show how such an ef\/fective Friedmann equation, (\ref{LQCcorr}), can be extracted from GFT condensates; they follow previous work in this GFT condensate setting, in which various derivations of ef\/fective ``Friedmann'' equations have been discussed \cite{LQCfromGFT,SteffenJHEP,NJP,PRL,JHEP}. These derivations use dif\/ferent approximations which need to be independently justif\/ied, and more work is needed on extracting such equations that in particular use the structure of the GFT interactions, which have been largely neglected so far.

Let us give some more detail on the recent results of \cite{VolumeDynamics,VolumeDynamicsShort}, which form the most complete derivation of such generalised Friedmann equations from GFT condensates so far. One focuses on the relational observable $V(\phi)$, giving the volume of a comoving part of the universe (the ``f\/iducial cell'') at a f\/ixed value of a scalar f\/ield $\phi$ that plays the role of relational clock. In classical general relativity, $V(\phi)$ satisf\/ies the equations
\begin{gather}\label{GRrelations}
\left(\frac{1}{3V}\frac{dV}{d\phi}\right)^2 = \frac{4\pi G_{\rm N}}{3},\qquad \frac{1}{V}\frac{d^2 V}{d\phi^2} = 12\pi G_{\rm N}.
\end{gather}
Note that the right-hand side is equal to a constant in both cases. The f\/irst equation gets modif\/ied in LQC where a term proportional to the matter density $\rho$ appears,
\begin{gather}\label{LQCcorr}
\left(\frac{1}{3V}\frac{dV}{d\phi}\right)^2 = \frac{4\pi G_{\rm N}}{3}\left(1-\frac{\rho}{\rho_c}\right),
\end{gather}
where $\rho_c$ is a critical (Planckian) density. In~\cite{VolumeDynamics}, the expectation values $V(\phi)$, $V'(\phi)$ and~$V''(\phi)$ are computed for a condensate state of the form (\ref{singlecon}), with an additional argument $\phi$ of the GFT f\/ield and the condensate wavefunction representing the physical scalar f\/ield; the mean f\/ield $\sigma(g_I,\phi)$ is restricted to a state of isotropic (equilateral) tetrahedra,
\begin{gather*}
\sigma(g_I,\phi)=\sum_{j=0}^\infty \sigma_j(\phi) {\bf D}^j(g_I),
\end{gather*}
where ${\bf D}^j(g_I)$ is an appropriate (f\/ixed) convolution of Wigner $D$-matrices with $\SU$ intertwiners implementing isotropy, such that the only geometric degree of freedom left in the condensate wavefunction $\sigma$ is the representation label~$j$. One then uses the equations of motion~(\ref{classicalEOM}) for an extension of the EPRL model~\cite{EPRL}; in terms of the variable~$j$ these take the form
\begin{gather}\label{jEOM}
A_j\partial_\phi^2\sigma_j(\phi)-B_j\sigma_j(\phi)+w_j\bar\sigma_j(\phi)^4=0,
\end{gather}
where $A_j$, $B_j$ and $w_j$ are parameters specif\/ied on one hand by the details of the GFT action and on the other hand by the intertwiners contained in the functions ${\bf D}^j(g_I)$, associated to the specif\/ic choices operated in the selection of the isotropic degrees of freedom. Note that~(\ref{jEOM}) decouples dif\/ferent $j$ components, which can hence be studied separately. This is a direct consequence of the particular form of the spin foam vertex chosen for the EPRL model, which possesses a specif\/ic pattern of identif\/ication of the spins. Other models, for instance the one proposed in~\cite{BaratinOritiHP}, will not necessarily have this decoupling property and therefore will lead to dif\/ferent evolution equations and, possibly, dif\/ferent physical predictions.

In the cases in which the decoupling of the dif\/ferent spins occurs, the system of equa\-tions~\eqref{jEOM} can be easily studied with conventional methods used for the study of classical dynamical systems: the problem boils down to determining the behaviour of a system of noninteracting particles in two dimensions (since the wavefunction is complex) subject to the action of an external potential, whose precise shape is determined by the GFT interaction term. Consequently, it is possible to determine, at least with qualitative methods, the properties of the full family of the solutions to the complete nonlinear equations, in the isotropic restriction, and to study their dependence on the parameters $A_j$, $B_j$ and $w_j$.

In the derivation of the analogue of (\ref{GRrelations}) and (\ref{LQCcorr}), one now uses (\ref{jEOM}) where interactions are neglected ($w_j\rightarrow 0$), so that the only relevant feature of the amplitude that is required is the plausible restriction that the spin labels of faces of the tetrahedra in the boundary of the four-simplex, when glued, are in fact the same. (\ref{jEOM}) includes a nontrivial kinetic operator $K$ due to the coupling to the scalar f\/ield, to get the corresponding Klein--Gordon kinetic term in the GFT amplitudes.

The main result of \cite{VolumeDynamics,VolumeDynamicsShort} is to show several cases in which, for appropriate choices of the couplings, the function $V(\phi)$ indeed satisf\/ies (\ref{GRrelations}), and to give the leading corrections. One f\/inds that the ef\/fective Newton's constant $G_{\rm N}$ is given in terms of the couplings $A_j$ and $B_j$, and hence originates in the structure of the GFT action. Interestingly, for a condensate in which all tetrahedra are constrained to have the same f\/ixed volume (given by a single spin $j_0$), exactly a~correction of the improved dynamics LQC form (\ref{LQCcorr}) is found, together with additional quantum corrections. This shows how GFT condensates can be governed by dynamics that are consistent not only with general relativity but also with LQC. Morever, the correct form of improved dynamics corrections arises without the need to introduce operators for the connection variables and argue for the dependence of holonomy corrections on $N$, as in~(\ref{piholo}), thus strengthening the connection between LQC improved dynamics and the more fundamental setting of GFT.

Several assumptions have to be made in order to obtain these results, see \cite{VolumeDynamics} for details. The most signif\/icant approximation is that, as in previous calculations, the GFT interaction term has to be neglected when matching the ef\/fective dynamics to (\ref{GRrelations}). As the magnitude of the mean f\/ield $\sigma(g_I)$ is related to the number of quanta, interactions become more and more relevant as the expectation value of $\widehat{N}(\phi_0)$ becomes large, so that this approximation becomes less justif\/ied. For the possible (polynomial) interaction terms that must be presumably added to the simplicial interaction term to take care of radiative corrections, higher monomials have a higher impact on the ef\/fective dynamics. As the coef\/f\/icients of these terms are not known, in absence of strong arguments from consistency of the quantum theory, we would be facing a problem of non-universality of the macroscopic dynamics. This is not a signal that GFT models do not reproduce general relativity in a large volume limit, but rather that the simple condensate state approximation (\ref{singlecon}) is not appropriate: as can be seen from the quantum equations of motion, as the number of tetrahedra increases, the coherent state deviates more and more from the exact state. The coherent state picture seems valid for a limited \emph{mesoscopic} regime in which there is the f\/irst formation of a GFT condensate. After the f\/irst stages, correlations between dif\/ferent quanta will not be negligible anymore and it will be necessary to replace simple coherent states with more complex states. This is not incompatible with the idea of GFT condensation, as these more complicated states can still be controlled by cosmological wavefunctions, and, e.g., be of the generalised form discussed in Section~\ref{gencond}. More work is needed to see whether general relativity can be recovered from the dynamics of these more complicated states.

As we mentioned in the introduction, nonlinear dynamical equations for quantum cosmology can also be motivated, independently of any assumptions on fundamental quantum gravity, as modelling inhomogeneities by interactions between dif\/ferent locally homogeneous patches. A~concrete model of this type was studied in~\cite{NonlinQC}, leading to a nonlinear, nonlocal extension of quantum cosmology; a more general class of models, serving as nonlinear extensions of LQC, had been given, with similar motivations, in~\cite{OldGFC}.

The construction of such models follows the general structure of building a quantum f\/ield theory in ``second quantisation'', where the kinetic operator in the quadratic term of the action corresponds to the wave equation satisf\/ied by a single-particle wavefunction. Starting from homogeneous, isotropic LQC def\/ined in terms of a volume variable $\nu$ and a massless scalar f\/ield~$\phi$, this leads to a general action of the form \cite{OldGFC}
\begin{gather}\label{LQCnonlinaction}
S[\Psi]=\half\sum_\nu \int_{\mathbb{R}} \dd\phi\, \Psi(\nu,\phi)\hat{\mathcal{K}}\Psi(\nu,\phi)+\sum_{j=2}^n\frac{\lambda_j}{j!}\sum_{\nu_1\ldots \nu_j}\int_{\mathbb{R}^j} \dd\phi\, f_j(\nu_i,\phi_i)\prod_{k=1}^j\Psi(\nu_k,\phi_k),
\end{gather}
where $\hat{\mathcal{K}}$ corresponds to the Hamiltonian constraint operator of a chosen LQC model,
\begin{gather*}
\hat{\mathcal{K}}\Psi(\nu,\phi):=-B(\nu)\big(\Theta+\partial_\phi^2\big)\Psi(\nu,\phi),
\end{gather*}
where $\Theta$ is a dif\/ference operator only acting on the variable $\nu$ (which takes discrete values, $\nu\in\nu_0\mathbb{Z}$) and $B(\nu)$ is a given function; for some examples from LQC, see~\cite{OldGFC}. The interaction kernels $f_j$ should be chosen to model inhomogeneities, as made more concrete in~\cite{NonlinQC}.

One could hope to obtain~(\ref{LQCnonlinaction}) or a similar form as an ef\/fective action for GFT condensates, where $\Psi$ either corresponds to a specif\/ic condensate wavefunction (for a homogeneous, isotropic universe) or to a perturbation of the GFT f\/ield around a background f\/ield, as proposed in a~similar context in~\cite{LOR}, depending on the interpretation of the ``second quantised'' scalar f\/ield~$\Psi$ in~(\ref{LQCnonlinaction}). Such a~result would show that a Hamiltonian constraint derived directly from general relativity captures the ef\/fective dynamics of GFT, and thus give evidence that the corresponding GFT model does indeed def\/ine a theory of quantum gravity. It is not necessary that such a~direct mapping should exist: what is really needed is evidence that the collective behaviour of many degrees of freedom, described by large-scale observables, behaves as predicted by general relativity. Obtaining such observables from the fundamental GFT variables presumably requires coarse-graining over many orders of magnitude (in the simplest case, rescaling by some power of~$N$), and it may not be possible to f\/ind an even approximate description of the macroscopic variables in terms of an ef\/fective ``LQC wavefunction''. Nevertheless, it is interesting to study the behaviour of nonlinear quantum cosmology models such as~(\ref{LQCnonlinaction}) in their own right, if only as toy models for the ef\/fective dynamics of GFT condensates. For instance, a linear but nonlocal extension of the Wheeler--DeWitt equation was recently studied in~\cite{Mairistudents}.

In order to extract cosmological phenomenology from the ef\/fective dynamics of GFT condensates, it is clearly necessary to have matter in the model. Depending on how realistic a~cosmology one wants to describe, this could for example be a radiation f\/luid, a scalar f\/ield modelling inf\/lation, or a massless scalar f\/ield that can serve as a relational clock~\cite{oldQCreview}. While most GFT models aim to def\/ine pure quantum gravity without matter, there are also models \cite{GFTmat1,GFTmat2} in which matter f\/ields are introduced explicitly, by enlarging the number of microscopic degrees of freedom associated to each basic building block. Alternatively, it has been suggested that matter degrees of freedom might already be included in the excitations of GFT degrees of freedom over a background~\cite{GFTemmat1,GFTemmat2} in which case there would be no need to add them exp\-li\-citly.

Both approaches can be followed in the cosmological setting of GFT condensates. The f\/irst approach leads to an extension of the domain space of the GFT f\/ield and the GFT action. This was f\/irst done in the context of the GFT condensates in \cite{JHEP}, where it was shown how the simplest extension to a scalar f\/ield led to the correct matter coupling to gravity for a~massless scalar f\/ield in the ef\/fective Friedmann equation. More recently, the more systematic constructions of~\cite{VolumeDynamics} give further support to this approach, by explicitly showing how the Friedmann equations of classical general relativity and their LQC corrections can arise from the dynamics of an appropriate condensate state and GFT model.

In the second approach, one tries to interpret corrections to an ef\/fective Friedmann equation (arising, for instance, from dif\/ferent types of GFT interactions) as matter f\/ields or as an ef\/fective cosmological constant. This possibility was outlined in \cite{NJP} where we point for further reference. It is potentially very interesting, as it may, for example, raise a hope to solve the cosmological constant problem through GFT condensates, but has not been explored much so far.

\section{Main results, open questions and future directions}\label{resultsec}

We close by summarising the main results, some of the remaining open questions and some directions for current and future research.

The f\/irst major achievement \cite{PRL,JHEP} was the development of a formalism for the def\/inition of group f\/ield theory condensates and for their geometric interpretation; the procedure outlined in Section~\ref{interpretsec} showed how an LQG ``graph'' consisting of a large number $N$ of disconnected elementary building blocks (open spin network vertices or gauge-invariant ``dipoles'') can be interpreted as a macroscopic, approximately smooth homogeneous geometry, if one imposes the criterion of wavefunction homogeneity that forces all building blocks to be in the same single-particle conf\/iguration. The def\/inition (\ref{metriccomp}) of an approximate continuum metric from geometric data in $N$-particle states was new and rather simple compared to other proposals in the literature. As we have described in Section~\ref{interpretsec}, some issues related to it need to be resolved, in particular as it only seems consistent in a frame in which the spatial metric is diagonal. The second main result of \cite{PRL,JHEP} was to show how an ef\/fective dynamics could be extracted from GFT Schwinger--Dyson equations, and reinterpreted as an ef\/fective quantum cosmology model. Agreement with the vacuum Friedmann equation of Lorentzian and Riemannian general relativity, in a WKB approximation, was found, and it was also shown how to include a massless scalar f\/ield, leading to the correct coupling of the scalar f\/ield to gravity in the ef\/fective Friedmann equation. Similar techniques were subsequently applied in \cite{LQCfromGFT} leading to a successful derivation of the LQC improved dynamics scheme from this GFT model coupled to matter.

Further work \cite{SteffenCQG,SteffenPRD,SteffenJHEP,NJP} clarif\/ied some of the conceptual and technical aspects of the framework. First, in \cite{SteffenCQG} analytical solutions to the ``Wheeler--DeWitt equation'' of~\cite{PRL,JHEP} were analysed in more detail and it was found that there are no WKB-like solutions satisfying the requirements of \cite{PRL,JHEP}: highly oscillating ``condensate wavefunctions'' do not correspond to nearly f\/lat tetrahedra and hence do not describe an approximate continuum (see Section~\ref{contlsec}). This already suggested that using the WKB approximation may not be justif\/ied. These results were followed up by \cite{NJP} where it was shown how the relation between cosmological observables and observables on the GFT Fock space involves a nontrivial scaling with the particle number~$N$ (see, e.g.,~(\ref{scalefdef})), which is not present in standard quantum cosmology. This allowed a~derivation of the form of LQC holonomy corrections in \cite{NJP}, depending on the unknown relation $N=N(a)$ between number of degrees of freedom $N$ and scale factor~$a$, similar to the lattice-ref\/inement picture of LQC \cite{BojOSIGMA,latticerefine}. It was also understood how the WKB approximation applied in the earlier works corresponds to a presumably unphysical regime in which individual ``atoms of space'' behave semiclassically; put dif\/ferently, it corresponds to a limit of $l_{{\rm Planck}}\rightarrow 0$ at f\/ixed~$N$, not the expected continuum limit which would be of double-scaling form $l_{{\rm Planck}}\rightarrow 0$ and $N\rightarrow\infty$. As a consequence, the ``Friedmann equations'' given in~\cite{NJP} contain additional terms compared to~\cite{PRL,JHEP} that come with extra powers of both $l_{{\rm Planck}}$ and $N$. It was suggested that these terms could act as ef\/fective matter components in the ef\/fective Friedmann equation.

Semiclassicality, a crucial issue in quantum cosmology where the emergence of a classical universe needs to be explained \cite{jeanluc}, was discussed in more detail in \cite{SteffenJHEP}: the mean-f\/ield approximation, encoded in the choice of state~(\ref{singlecon}) and a large number of quanta~$N$, already corresponds to a semiclassical limit; the mean-f\/ield conf\/iguration $\sigma(g_I)$ represents a classical GFT f\/ield conf\/iguration rather than a f\/irst-quantised quantum cosmology wavefunction. Consequently, it should be interpreted as def\/ining a ``patch density'' on a minisuperspace of geometric data, given by the parallel transports $g_I$ or area bivectors $B_I$ -- the hydrodynamic description of a~GFT condensate corresponds to a~``f\/luid'' on such a minisuperspace, or a statistical distribution of patches with locally dif\/ferent geometries~\cite{SteffenJHEP}. This interpretation implements a quantum-to-classical transition exactly of the type needed in inf\/lation, where the quantum f\/luctuations of the inf\/laton f\/ield over a homogeneous, isotropic vacuum state lead to classical statistical inhomogeneities and anisotropies. A~conceptually rather similar idea, akin to the hydrodynamic picture of a Bose--Einstein condensate, was in fact recently proposed in~\cite{stephonjoao} to solve this ``measurement problem'' of inf\/lation. The formalism of GFT condensates here connects directly to some of the main conceptual questions in inf\/lation, and suggests that inhomogeneities can arise from a non-zero spread of the density prof\/ile on minisuperspace~\cite{SteffenJHEP}.

The conceptual basis of the framework, and the map between the fundamental GFT dynamics and the ef\/fective description in terms of cosmology, are now rather well understood. What remains is to apply these ideas to concrete GFT models of interest, constructing appropriate, physically interesting condensate states. So far, it has proven rather dif\/f\/icult to include the GFT interaction terms into the analysis, and indeed almost all of the existing work has neglected them, either in a weak-coupling limit or by choosing specif\/ic states and Schwinger--Dyson equations that are unaf\/fected by the interactions. This is ultimately unsatisfactory, as results obtained from such calculations would not be sensitive to most of the details of the GFT or spin foam model used, and the approximation one makes appears not to be justif\/ied as the number of GFT quanta becomes large. In the construction of states, recently \cite{GeneralCondensates, HorizonEntropy} more emphasis has been put on constructing a more general class of condensate states that allow the use of a nontrivial LQG graph structure and its associated topology, and also generalise the setting from spatial homogeneity to spherical symmetry, opening an avenue for exploring black holes and their thermodynamics: indeed it has been shown \cite{HorizonEntropy} that an appropriately def\/ined horizon entanglement entropy for such generalised condensates gives an area law with subleading logarithmic corrections. The cosmological phenomenology of these generalised condensates has not been explored much, as the resulting ef\/fective dynamical equations will be more involved, and additional approximation techniques presumably need to be developed to handle them.

A main question that this approach faces is whether its ef\/fective dynamics, at some semiclassical level, can reduce to the ef\/fective dynamics of LQC, or more generally to general relativity up to corrections at high curvature. Here, we can list the kinematical results of \cite{NJP} that derived the general form of holonomy corrections as functions of $N$, and the more concrete calculations of \cite{LQCfromGFT,SteffenJHEP} that gave evidence for the {\em improved dynamics} form of these holonomy corrections. A derivation of LQC inverse-triad corrections \cite{LQCrev1,LQCrev2} would presumably require a clearer understanding of the def\/inition of inverse volume operators in the GFT setting. Dynamically, the clearest and promising results that show a derivation of modif\/ied Friedmann equations, of the precise LQC form, from the dynamics of GFT condensates have been obtained in \cite{VolumeDynamics,VolumeDynamicsShort}, assuming a certain matter coupling in GFT, a type of semiclassical limit, and neglecting GFT interactions (see Section~\ref{LQCrelation} for further discussion). The latest results in particular show that such a derivation of LQC dynamics is possible and may be rather universal, given that only specif\/ication of the kinetic term seems required. They allow explicit contact with LQC and beyond, as one can also compute interesting modif\/ications to general relativity and LQC, which characterise dif\/ferent GFT models and allow further distinguishing between them.

Further progress will depend on the development of better approximation techniques, in order to handle more complicated condensate states, and include more information about the GFT dynamics. Numerical techniques, which have so far hardly been used (see, however, \cite{Pithis}), will become useful in the more detailed study of various GFT models. Apart from this, most other open questions address the validity of the GFT condensate picture itself. The approximations implied by assuming that a certain type of condensate state is a good approximation to a physical state of the full theory need to be quantif\/ied and verif\/ied in detail. To some extent, one can rely on insight from related situations in condensed matter physics, where the types of states we have considered describe a weakly interacting, diluted Bose--Einstein condensate, and one can see where this approximation breaks down by looking at the ef\/fect of interactions and by computing correlation functions \cite{BECbook}. In the GFT context, similar calculations should be performed, which connect to results on the phase structure and renormalisation of GFT models (see Section~\ref{renormsec}). As we have said, one would like to understand the breakdown of the condensate approximation in terms of a phase transition, and relate it to the scenario of ``geometrogenesis'' where the geometric phase given by a GFT condensate was preceded by a~dif\/ferent, non-geometric phase.

Again on a more technical level, the reconstruction procedure of Section~\ref{interpretsec} needs to be ref\/ined and understood better in the setting of simplicial geometry and LQG; more generally, a~clearer def\/inition of isotropic and anisotropic geometries, in terms of a detailed mapping to the possible Bianchi geometries, needs to be developed, in order to extend the existing calculations for isotropic models to anisotropies. For inhomogeneities, in Section~ \ref{interpretsec} we have mentioned one way of identifying them in the mean-f\/ield approximation of the simplest condensates, but there may be clearer ways of including inhomogeneities, allowing their localisation on the emergent space given by the GFT condensate. Indeed, the main obstacle to a derivation of a power spectrum of inhomogeneities in a GFT condensate seems to be the absence of a good notion of ``wavenumber'' (usually def\/ined through the spectrum of a suitable Laplace operator). The existing results of \cite{SteffenPRD,SteffenJHEP} are not fully conclusive, since they only describe homogeneous perturbations or global quantities such as (\ref{integral}) characterising the inhomogeneities. The ideas of~\cite{JHEP} for including inhomogeneities as a type of phonons, while quite natural from the condensate perspective, have not been explored in much detail. In many ways, a clearer treatment of inhomogeneities, both technically and conceptually, in GFT condensates is one of the key open issues, given that the availability of detailed observations of the CMB power spectrum would of\/fer a very concrete way to put a whole class of quantum gravity models to a direct observational test.

\looseness=-1 A general problem faced by the GFT/LQG community in order to obtain more realistic mo\-dels is how matter is to be included into models of interest. GFT condensates allow approaching it in the same way as for pure gravity models: given a proposal for including matter, one can def\/ine the appropriate condensate states, compute the ef\/fective dynamics, and check that they reduce to the classical Friedmann equation coupled to matter, perhaps up to quantum corrections. We have shown one such calculation, in the EPRL model coupled to a massless scalar f\/ield, in Section~\ref{LQCrelation}, and the promising results should motivate further work on similar models.

To conclude, group f\/ield theory condensates have already been shown to provide very ge\-ne\-rally applicable methods for extracting an ef\/fective description of the collective behaviour of a~large number of degrees of freedom of quantum geometry, def\/ined by a GFT. Techniques and approximation methods for obtaining and solving ef\/fective equations can, to some extent, be borrowed from condensed matter physics, but need to be modif\/ied to the peculiar structure (in particular, the combinatorial type of nonlocality) of GFT models. Developing such methods further seems to be the main step missing for a full derivation of cosmological phenomenology of GFT, including GFT interactions. Several results, in particular those of~\cite{VolumeDynamics,VolumeDynamicsShort}, already indicate a close relation of the resulting semiclassical dynamics to ef\/fective equations in LQC.

\subsection*{Acknowledgements}

We thank Martin Bojowald, Daniele Oriti, Andreas Pithis and Edward Wilson-Ewing and the anonymous referees for helpful comments on the manuscript. The research leading to these results has received funding from the People Programme (Marie Curie Actions) of the European Union's Seventh Framework Programme (FP7/2007-2013) under REA grant agreement n$^{\rm o}$~622339 and from the John Templeton Foundation through the grant PS-GRAV/1401.

\pdfbookmark[1]{References}{ref}
\LastPageEnding

\end{document}